\documentclass[oldversion]{aa}
\usepackage{graphicx}
\usepackage[varg]{txfonts}
\usepackage{natbib}
\usepackage[normalem]{ulem}
\renewcommand{\tabcolsep}{0.7mm}
\bibpunct{(}{)}{;}{a}{}{,} % to follow the A&A style

\def\ecs{erg~cm$^{-2}$s$^{-1}$}

\usepackage{color}

\begin{document}

\title{Neutron star cooling and the rp process \\ in thermonuclear
  X-ray bursts\thanks{Table~\ref{tabj} is only available in
    electronic form at the CDS via anonymous to cdsarc.u-strasbg.fr or
    via http://cdsweb.u-strasbg.fr/cgi-bin/gcat?/A+A/}}
\titlerunning{Neutron star cooling and the rp process in thermonuclear
  X-ray bursts}

\authorrunning{in 't Zand et al.}

\author{J.J.M.~in~'t~Zand\inst{1}, M.E.B. Visser\inst{1,2},
  D.K. Galloway\inst{3,4}, J. Chenevez\inst{5}, L. Keek\inst{6,7},\\
  E. Kuulkers\inst{8}, C. S\'{a}nchez-Fern\'{a}ndez\inst{8},
  H. W\"{o}rpel\inst{9}}

\institute{     SRON Netherlands Institute for Space Research, Sorbonnelaan 2,
                3584 CA Utrecht, the Netherlands; {\tt jeanz@sron.nl}
                \and
                University College, Utrecht, the Netherlands
                \and
                School of Physics and Astronomy, Monash University, VIC 3800, Australia
                \and
                Monash Center
                for Astrophysics, Monash University, VIC 3800, Australia
                \and
                National Space Institute, Technical University
                of Denmark, Elektrovej 327-328, DK-2800 Lyngby, Denmark 
                \and
                CRESST and X-ray Astrophysics Laboratory NASA/GSFC,
                Greenbelt, MD 20771, USA
                \and
                Department of Astronomy, University of Maryland, College
                Park, MD 20742, USA
                \and
                European Space Astronomy Centre (ESA/ESAC),
                Science Operations Department, 28691 Villanueva de la
                Ca\~{n}ada, Madrid, Spain 
                \and
                Leibniz-Institut f\"{u}r Astrophysik Potsdam, An der
                Sternwarte 16, 14482 Potsdam, Germany
         }

\abstract{When the upper layer of an accreting neutron star
  experiences a thermonuclear runaway of helium and hydrogen, it
  exhibits an X-ray burst of a few keV with a cool-down phase of
  typically 1~minute. When there is a surplus of hydrogen, hydrogen
  fusion is expected to simmer during that same minute due to the rp
  process, which consists of rapid proton captures and slow
  $\beta$-decays of proton-rich isotopes. We have analyzed the
  high-quality light curves of 1254 \typeout{NUMBERS} X-ray bursts,
  obtained with the Proportional Counter Array on the Rossi X-ray
  Timing Explorer between 1996 and 2012, to systematically study the
  cooling and rp process. This is a follow-up of a study on a
  selection of 37 bursts from systems that lack hydrogen and show only
  cooling during the bursts. We find that the bolometric light curves
  are well described by the combination of a power law and a one-sided
  Gaussian. The power-law decay index is between 1.3 and 2.1 and
  similar to that for the 37-bursts sample. There are individual
  bursters with a narrower range.  The Gaussian is detected in half of
  all bursts, with a typical standard deviation of 50~s and a fluence
  ranging up to 60\% of the total fluence. The Gaussian appears
  consistent with being due to the rp process. The Gaussian fluence
  fraction suggests that the layer where the rp process is active is
  underabundant in H by a factor of at least five with respect to
  cosmic abundances.  Ninety-four percent of all bursts from
  ultracompact X-ray binaries lack the Gaussian component, and the
  remaining 6\% are marginal detections. This is consistent with a
  hydrogen deficiency in these binaries. We find no clear correlation
  between the power law and Gaussian light-curve components.}


\maketitle

\section{Introduction}
\label{sec:intro}

The Galaxy is host to at least 100 neutron stars that accrete hydrogen
and/or helium from a Roche-lobe overflowing companion star.  The
hydrogen and helium can burn in an unstable fashion during
thermonuclear shell flashes \citep[for reviews,
  see][]{lewin1993,stroh2006,galloway2008}. The typical duration is
one minute, and the typical frequency is once every few hours. The
neutron star photosphere reaches typical temperatures of $\sim10^7$~K,
and the thermonuclear runaway expresses itself as a bright X-ray
burst, easily detectable from anywhere in the Galaxy.

Thermonuclear X-ray bursts are powered by mainly four nuclear reaction
chains \citep[e.g.,][]{fujimoto1981,bildsten1998,jose2010}:
\newcounter{ll}
\begin{list}{\arabic{ll}.}{\leftmargin=0.4cm \parskip=0cm \partopsep=0cm \itemsep=0cm \parsep=0cm \topsep=0.cm \usecounter{ll}}
\item the CNO cycle \citep{taam1979}, acting on hydrogen forming
  helium and catalyzed by CNO. The net reaction captures four protons
  and emits one $\alpha$ particle and two positrons, neutrinos, and
  photons. It includes two $\beta$ decays. It may be responsible for
  ignition at the lowest accretion rates; see \cite{fujimoto1981} and
  \cite{peng2007};
\item the 3$\alpha$ process \citep{joss1978}, acting on helium forming
  carbon. It is usually, if not always, responsible for the ignition
  \citep[i.e., for mass accretion rates higher than 1\% of the
    Eddington limit;][]{fujimoto1981};
\item the $\alpha$p process, acting on products of the previous two
  chains forming heavier elements. It occurs when the temperature is
  high enough ($>5\times10^8$~K) and results from a breakout from the
  CNO cycle through $\alpha$-captures by $^{15}$O;
\item the rp process, acting on products of the previous chains and
  any hydrogen that is left from the previous chains
  \citep{wallace1981,schatz2001}. The higher the temperature, the
  longer the chain of the rp process because the protons need to
  overcome increasingly higher Coulomb barriers of heavier
  isotopes. Some branches of the rp process are slow. Although the
  proton capture rates are high, the $\beta$ decays are sometimes much
  slower and constitute bottlenecks in the chain. These are
  responsible for late nuclear burning in X-ray bursts. Notorious
  waiting points are at $^{21}$Mg, $^{33}$Ar, and $^{48}$Mn
  \citep[e.g.,][see also Appendix~\ref{appa}]{fisker2008} with decay
  half-lives of up to tens of seconds. Reaction rates in the rp
  process are sometimes ill-constrained experimentally. The relevance
  of this for the time profiles of the burst luminosity has recently
  been assessed by \citet[see also
    \citealt{woosley2004}]{cyburt2016}. They found that the
  uncertainties in the rates of about ten rp and $\alpha$p reactions
  introduce significant uncertainty in the time profiles.
\end{list}

\noindent
The existence of two types of fuel that can burn independently from
each other, hydrogen and helium, results in the so-called burning
regimes. Through the CNO cycle, hydrogen ignites at lower temperatures
and pressures than helium. For $T<8\times10^7$~K, the nuclear power of
the CNO cycle increases faster with $T$ than the cooling, and a
runaway occurs. For higher temperatures, the nuclear power
$T$-dependence levels off to the radiative cooling dependence and the
burning becomes continuous.  When the burning of a mass parcel of
hydrogen is faster than the time for it to reach helium ignition
conditions, a thermonuclear flash ignites in a
helium-rich/hydrogen-poor environment. At higher accretion rates,
ignition conditions are reached faster than it takes for stable
hydrogen burning, and the flash occurs in a layer containing hydrogen
and helium. This results in three burning regimes
\citep{fujimoto1981}: mixed hydrogen-helium bursts ignited by hydrogen
at low accretion rates (regime 3), pure helium bursts at medium
accretion rates \citep[regime 2; between 3\%\ and 6\% of Eddington for
  $Z_{\rm CNO}=0.01$;][]{bildsten1998}, and mixed hydrogen-helium
ignited by helium at high accretion rates (regime 1). At the highest
accretion rates, helium burning may become partially stable
\citep[][]{zand2004,keek2016}.

\begin{table}
\begin{center}
\caption[]{Sample of 2288 bursts detected with RXTE-PCA from 60
  sources (a) and of the selection of 1254 bursts employed in the current
  study (b).\label{tab1}}
\begin{tabular}{lrr|lrr}
\hline\hline
Source & \multicolumn{2}{c|}{Numbers} & Source & \multicolumn{2}{c}{Numbers} \\
 & a & b & & a & b \\
\hline
4U 0513-40  &  20  & 17 & GX 3+1           & 3  &  3 \\
4U 0614+09  &   1  &  0 & EXO 1745-248     & 22 & 16 \\
EXO 0748-676& 160  &108 & 4U 1746-37       & 29 & 13 \\
4U 0836-429 &  17  &  1 & IGR J17473-2721  & 45 & 40 \\
2S 0918-549 &   4  &  1 & GRS 1747-312     &  7 &  2 \\
4U 1254-69  &   7  &  4 & SAX J1747.0-2853 & 27 & 15 \\
4U 1323-62  &  40  & 25 & SAX J1748.9-2021 & 29 & 29 \\
Cir X-1     &  13  &  0 & IGR J17480-2446  &303 & 15 \\
4U 1608-52  &  56  & 43 & IGR J17498-2921  &  2 &  0 \\
4U 1636-536 & 388  &303 & SAX J1750.8-2900 &  7 &  5 \\
MXB 1658-298&  26  &  8 & IGR J17511-3057  & 10 &  9 \\
XTE J1701-462&  3  &  3 & IGR J17597-2201  &  9 &  9 \\
4U 1702-429 &  51  & 41 & SAX J1806.5-2215 &  4 &  4 \\
4U 1705-44  &  94  & 64 & SAX J1808.4-3658 &  9 &  7 \\
XTE J1709-267&  3  &  1 & XTE J1810-189    &  9 &  3 \\
XTE J1710-281 & 46 &  9 & SAX J1810.8-2609 &  6 &  6 \\
IGR J17191-2821& 5 &  5 & XTE J1812-182    &  7 &  0 \\
XTE J1723-376 &  3 &  0 & XTE J1814-338    & 28 & 25 \\ 
4U 1722-30 &  4    &  3 & GX 17+2          & 15 &  3 \\ 
4U 1728-34 & 175   &137 & 4U 1820-303      & 16 & 15 \\
MXB 1730-335 & 127 & 45 & GS 1826-24       & 78 & 58 \\
KS 1731-260 & 27   & 25 & XB 1832-330      &  1 &  0 \\
SLX 1735-269& 1    &  1 & Ser X-1          & 19 & 13 \\
4U 1735-444 & 23   & 17 & 4U 1850-086      &  1 &  0 \\
XTE J1739-285 & 6  &  6 & HETE J1900.1-2455&  8 &  8 \\
KS 1741-293 & 1    &  0 & Aql X-1          & 75 & 53 \\
GRS 1741.9-2853 & 8&  0 & XB 1916-053      & 14 & 10 \\
1A 1742-294 & 87   & 5 & XTE J2123-058    &  6 &  3 \\
SLX 1744-299/300&24& 14 & 4U 2129+12       &  6 &  1 \\
1A 1744-361 & 3    &  3 & Cyg X-2          & 70 &  0 \\
\hline\hline
\end{tabular}
\end{center}
\end{table}

For a subset of binaries, the so-called ultracompact X-ray binaries
\cite[UCXBs, with orbital periods shorter than 80 min;][]{nelson1986},
there is a deficiency of accreted hydrogen and the nuclear reactions
simplify and become faster. The lack of hydrogen precludes a strong rp
process and prolonged burning that would be visible as shoulders in
type I X-ray burst decays. Possibly as many as half of all X-ray
bursters fall in this category \citep{zand2007}. However, the burst
rate in these systems is usually rather low (i.e., one every few days
or even longer gaps). Most X-ray bursts therefore come from
hydrogen-rich systems, and it is expected that all four nuclear
reaction chains mentioned at the start of this section occur.

Thermonuclear X-ray bursts were first detected in 1969
\citep{belian1972,kuulkers2009}. The number of detections grew into
the hundreds in the mid- to late-1970s
\citep[e.g.,][]{grindlay1976,lewin1993} and into the thousands with
the advent of X-ray telescopes with large fields of view in the 1990s
and onward
\citep[e.g.,][]{zand2004a,nakagawa2004,chelovekov2011,jenke2016} or
narrow-field telescopes with observation programs that emphasized
accreting neutron stars. The best example of the latter is the
high-throughput Proportional Counter Array (PCA) on the Rossi X-ray
Timing Explorer (RXTE), which was operational between 1996 and
2012. It detected more than 2200 thermonuclear X-ray bursts with a
high throughput \citep{galloway2008,galloway2010b}.

\cite{zand2014a} analyzed in detail the light curves of a subset of 37
X-ray bursts that were detected with the PCA. The down selection from
more than 2200 to 37 was made in two steps: 1) requiring that the
decay, as judged by eye, is smooth, and 2) requiring that the covered
dynamic range in flux is wide (more than about 50). This naturally
selects hydrogen-poor bursts and UCXBs. The most important result of
these authors is that the decay in photon count rate and energy flux
can best be modeled by a simple power law (in contrast to the commonly
employed exponential decay function), and that the power-law decay
index for the energy flux is on average 1.8, with a broad range
between 1.3 and 2.4. The most common index of 1.8 is consistent with
the electrons carrying most of the heat capacity of the cooling
gas. Each burst is consistent with a single power law, which is at
odds with the theory presented in this study.

\begin{figure}[t] % from t to p for referee print version
\begin{center}
\includegraphics[width=\columnwidth,angle=0,trim=1.5cm 7.5cm 1.5cm 6cm]{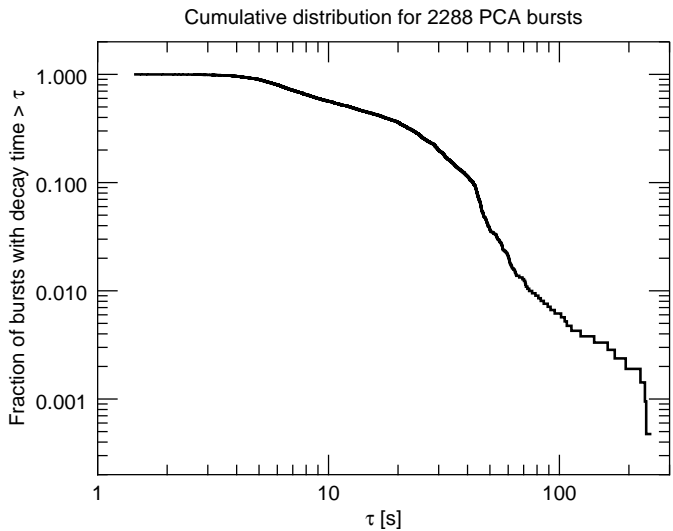}
\end{center}
\caption{Cumulative distribution of exponential decay time $\tau$, as
  found from fitting Eq.~\ref{eqn1} to the photon count rate
  decay. The employed light curves have 1~s time resolution and
  concern all RXTE-PCA bursts, except for 164 bursts with insufficient
  data coverage or almost unconstrained decay times. The most
  frequent decay time is 5 s. The average is 17.7 s.\label{fig2}}
\end{figure}

This study is a follow-up of this light curve study of 37 bursts, in
which we do not make a selection based on a smooth decay. Therefore,
it includes many hydrogen-rich bursts. This 1) provides us with a data
set to study the rp process, which because of the waiting points
prolongs into the cooling phase, and 2) allows us to study the cooling
over many more bursts than 37 after we separate its contribution in
the burst decay from that of the rp process. The data set presented by
the PCA is the basis of our study. It is the best data set available
on X-ray bursts because it is large and provides a wide dynamic range
in photon count rates.

In Sects.~\ref{secdata} and \ref{secspectral} we introduce the data of
our study and explain how the time histories of the bolometric flux
were extracted for each burst. In Sect.~\ref{secanalysis} we present
our approach to modeling the time histories, and in
Sect.~\ref{results} the results from that modeling, illustrated with
histograms and diagrams. In Sect.~\ref{secdiscussion} we discuss these
results, while in Sect.~\ref{secconclusion} we present the conclusions
of our study.

\section{Data}
\label{secdata}

\subsection{Data overview}

The PCA, operational from January 1996 to January 2012, consisted of a
linear array of five proportional counter units (PCUs) with a total
photon-collecting area of 8000~cm$^2$, a bandpass of 2 to 60 keV, and
a spectral resolution of about 17\% at 6 keV \citep{jahoda2006}. Each
PCU had two proportional counter chambers on top of each other: a top
propane layer, and a bottom xenon layer. The xenon layer is the main
instrument. The propane layer was used as an anticoincidence counter,
although it did occasionally provide scientific value since it
extended the bandpass to somewhat lower energies
\citep[e.g.,][]{keek2012b}. Observations were made with various
combinations of PCUs. In general, the average number of active PCUs
decreased from five early in the mission to one at the end. The center
PCU (number 2, counting from 0) was almost always operational. The PCA
electronics typically needed 10~$\mu$s to process a single event. For
event rates typical for type I bursts (10$^{4}$~s$^{-1}$), the live
time fraction of the PCA was affected by a few percent.

The PCA could be simultaneously read out by six event analyzers (EAs)
that could be programmed in any of seven basic read-out modes. One EA
always employd the standard-1 mode, yielding photon count rates at
0.125 s resolution for each PCU separately and no photon energy
resolution. An often-used mode was the science event mode, which
provided the energy information (usually in 64 channels) and timing
information (often at 125 $\mu$s resolution) for every detected
event. Another often-used mode is the good-xenon mode, which provided
0.95 $\mu$s time resolution and 256-channel spectra. It was only
useful for faint bursts because it more easily overflowed the
telemetry than the science event mode. We employed the science event
mode because we required an energy resolution that was capable of
determining the bolometric flux, and in incidental cases we used the
good-xenon mode.

The list of thermonuclear X-ray bursts that RXTE detected was obtained
from the Multi-INstrument Burst ARchive
\citep[MINBAR;][]{galloway2010b}. In addition to RXTE/PCA data, MINBAR
contains the bursts detected with BeppoSAX/WFC \citep{jager1997} and
the still operational INTEGRAL/JEM-X \citep{lund2003}. The PCA list in
MINBAR consists of 2288 \typeout{NUMBERS} bursts from 60
\typeout{NUMBERS} sources (i.e., this is slightly more than half the
currently known burster population). Some sources only exhibited one
burst in the PCA (e.g., KS 1741-293), while others had close to 400
(e.g., 4U 1636-536). Table~\ref{tab1} lists the burst counts per
source (column a).

For a broad perspective of the burst sample and an easy comparison
with burst parametrizations elsewhere, we show in Fig.~\ref{fig2} the
cumulative distribution of the exponential decay time $\tau$ as
derived from observed photon rates for all bursts. We included all bad
fits since we are only interested in a general picture of timescales
and did not draw any further conclusions from these numbers. The most
common decay time is 5~s, 50\%\typeout{NUMBERS} of all bursts have
decay times shorter than 10 s, and 1\% have decay times longer than 70
s. The average is 18.9 s.

\subsection{Data selection}

Bursts are often incompletely covered or have signal-to-noise ratios
that are too low on the peak fluxes to meaningfully study the decay
(i.e., they yield parameter values with large and, thus,
indiscriminate uncertainties). We selected only bursts with a
continuous data stretch that included the burst start and bursts with
relative Poisson errors (after background subtraction) during burst
peak better than 5\% for one-second exposures. This decreased our
sample by roughly 45\% to 1254\typeout{NUMBERS}. The number of bursts
left per source is specified in Table~\ref{tab1} (column b). Much of
the decrease in the sample is due to the many faint bursts on top of a
bright persistent flux from IGR~J17480-2446
\citep[e.g.,][]{linares2012}, MXB 1730-335 \citep{bag13}, 1A 1742-294
\citep{galloway2008}, Cyg X-2 \citep[e.g.,][]{smale1998} and GX 17+2
\citep[e.g.,][]{kuu02b}. We note that our analysis is biased toward
bursts from 4U 1636-536, 4U 1728-34, GS 1826-24, and Aql X-1 because
they have the most bursts and the highest signal-to-noise ratio.

\section{Spectral analysis}
\label{secspectral}

\subsection{Approach}
\label{sub:approach}

Spectra were extracted in fine enough time bins (from 1 s early on in
the burst to typically 16 s at the end of the burst) up to mostly 300
s after the burst start time. This 300 s time limit is longer than
employed in \cite{galloway2008}, which only covers the brightest
portion of the burst (often 30 s and at most out to 150-200 s). The
burst start time was determined from the time history of the photon
count rate at 1/8~s time resolution; for further details, we refer to
the upcoming MINBAR catalog paper (Galloway et al., in
prep.). Furthermore, we extracted for each burst a pre-burst spectrum
from data taken in the same EA read-out mode and between 100 and 16 s
before the burst start, except for a few tens of cases when the data
start later than 100 s before the burst start, but before the 16 s
mark. All spectra were rebinned such that each spectral bin contained
at least 15 photons to ensure applicability of $\chi^2$ as the
goodness-of-fit parameter. Corrections were applied for the instrument
dead time following the prescription from the instrument
team\footnote{{\tt http://heasarc.gsfc.nasa.gov/..\newline
    ../docs/xte/recipes/pca\_deadtime.html}}.

\begin{figure}[t]
\begin{center}
\includegraphics[width=\columnwidth,angle=0,trim=2.5cm 2cm 7.5cm
  10.8cm]{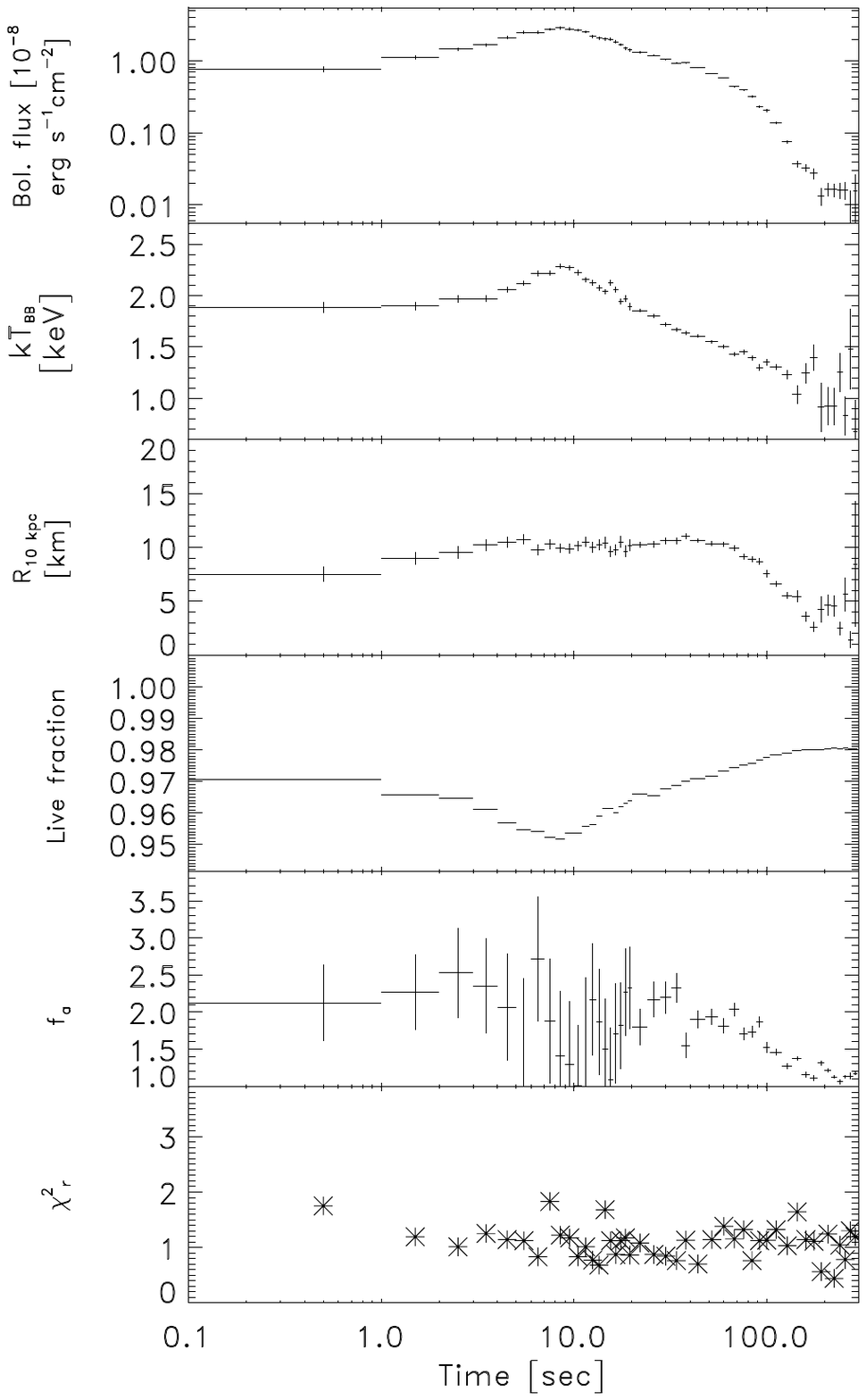}
\end{center}
\caption{Example of time-resolved spectroscopy of a burst for a burst
  from GS 1826-24 that was detected at MJD~51811.750. The typical
  number of degrees of freedom in these spectral fits is 20. The
  panels from top to bottom show the time history of the bolometric
  flux of the best-fit model blackbody, its temperature, its
  normalization in terms of radius in km of emission sphere when
  located at 10 kpc distance, the fraction of the time that the
  detector is susceptible to photon detection (allowing for the
  detector dead time, see text), the best-fit $f_{\rm a}$ factor, and
  the reduced chi-squared of the fit.
  \label{figexamplesp}}
\end{figure}

\begin{figure}[t]
\begin{center}
\includegraphics[width=\columnwidth,angle=0,trim=1.5cm 2cm 1.5cm
  0cm]{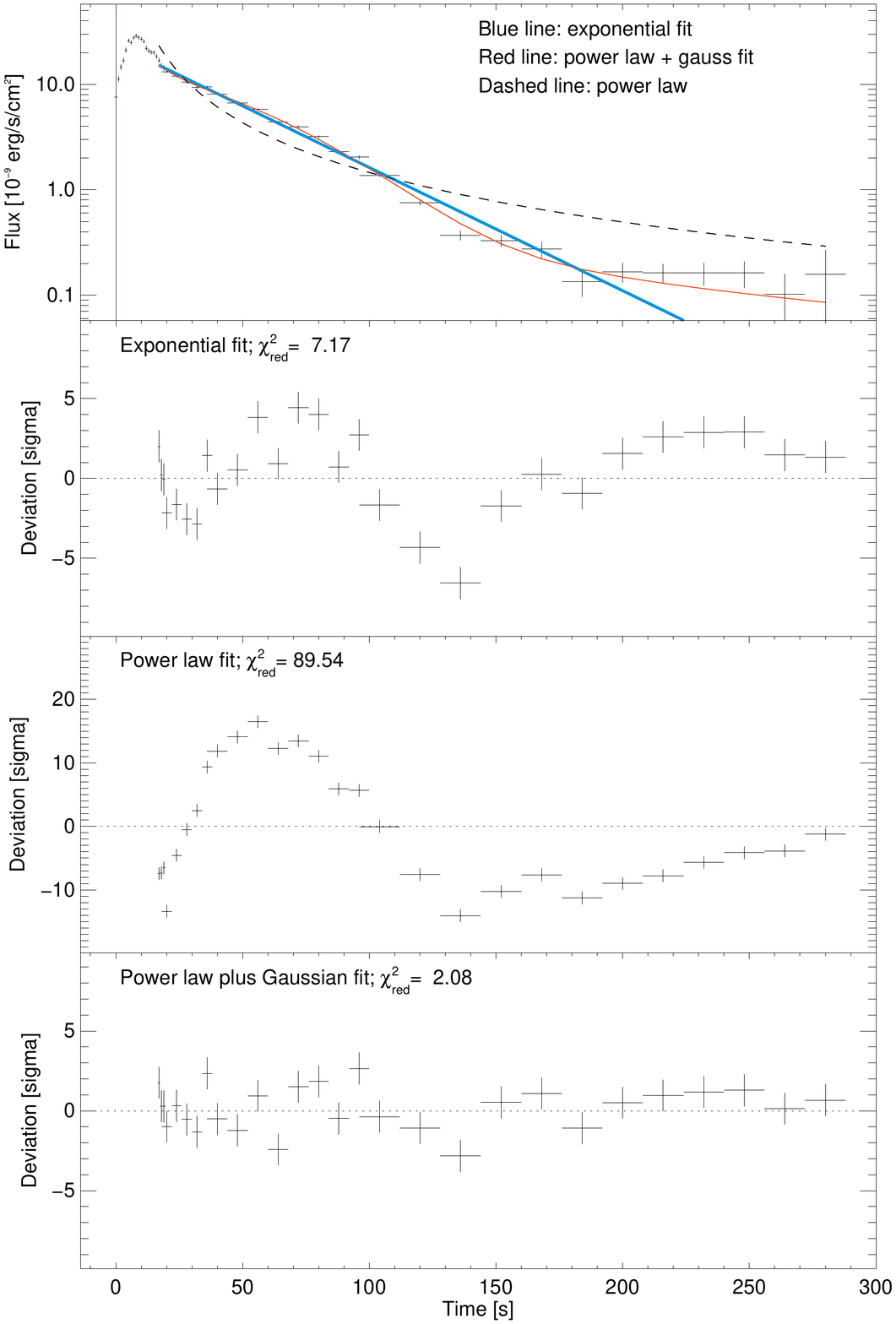}
\end{center}
\caption{Example of a fit to the decay phase of a bolometric light
  curve of a burst. This is the same burst as in
  Fig.~\ref{figexamplesp}. The top panel shows in logarithmic scale
  the bolometric flux and the best-fit models for an exponential decay
  (dashed curve; Eq.~\ref{eqn1}), the power law (blue curve;
  Eq.~\ref{eqn2}), and the power-law plus Gaussian function (red
  curve; Eq.~\ref{eqn3}). The second, third, and fourth panels show
  the residuals with respect to models following Eqs.~\ref{eqn1},
  \ref{eqn2}, and \ref{eqn3}. The power-law plus Gaussian function is
  the best model. \label{figexample}}
\end{figure}

The extracted spectra encompass all emission within the field of view
and are expected to contain the following components: burst radiation,
cosmic X-ray background, particle-induced background, emission from
other sources in the field of view, and the non-burst flux from the
burst source itself (due to the accretion process). The pre-burst
spectrum is considered as one combined measurement of all the
components except for the burst radiation. We modeled it through a
disk blackbody \citep{mitsuda1984} plus power law with the inner disk
temperature, photon index, and normalizations of both components as
free parameters. Each burst spectrum was modeled by a combination of
this pre-burst model and a Planck function for the burst radiation
with effective temperature and normalization as free parameters.

It has recently been found that the non-burst accretion flux changes
roughly in tandem with the burst flux. \cite{zand2013} analyzed the
broadband Chandra-PCA spectrum between 0.8 and 20 keV of a very bright
burst from SAX J1808.4-3658 and discovered that the spectrum could be
satisfactorily modeled if the amplitude of the pre-burst spectrum
during the burst was left free to change by a factor of $f_{\rm a}$
while keeping its shape constant. $f_{\rm a}$ was measured to peak at
a value of about 30. \cite{worpel2013,worpel2015} repeated this
exercise on all PCA bursts (approximately the same data set as
discussed in the present paper) and found this to be true in general.
$f_{\rm a}$ ranges between 1 and a few hundred during the burst
peak. When the burst decays, $f_{\rm a}$ generally decreases to
approximately unity. There are two interpretations of this '$f_{\rm
  a}$ effect': the accretion rate increases in tandem with the burst
because of the Poynting-Robertson effect \citep{worpel2013,worpel2015}
or a fraction of the burst photons is Compton scattered by an
optically thin medium (possibly the accretion disk corona) into the
line of sight \citep{zand2013,keek2014}. Support for scattering and
cooling of the corona comes also from measurements at photon energies
above 30 keV \citep[e.g.,][]{maccarone2003,ji2015,kajava2017a}.

This $f_{\rm a}$ effect implies that an important choice has to be
made in the spectral modeling of the bursts: leave $f_{\rm a}$ free,
or keep it fixed at $f_{\rm a}=1$. The latter constitutes the
traditional method of modeling burst spectra and assumes that the
accretion emission is unaffected by the bursts. We chose to perform a
full analysis with both methods, showing the results for a free
$f_{\rm a}$ in the main part of this paper and those of the fixed
$f_{\rm a}=1$ in Appendix~\ref{appb}. We find that the population-wide
perspective of the results is the same, but results on individual
sources may differ somewhat.

We applied the $f_{\rm a}$ method in a somewhat different manner than
\cite{worpel2013,worpel2015}. First, we applied a longer exposure time
for the pre-burst spectrum of usually 84 s versus 16 s by
\cite{worpel2013,worpel2015}. Second, we fit the {\it \textup{same}}
spectral model to {\it \textup{all}} pre-burst spectra, while Worpel
et al. fit different models to different bursts. The combination of a
disk blackbody and a power law was found to be most generally
applicable to all pre-burst spectra, with 63(90)\% of all spectra
having $\chi^2_\nu<2(4),$ and all burst spectra were finally
acceptable. We stress that we employed this model purely empirically
and ignored any physical interpretation of it. Third, we did not
distinguish between non-burst emission from the burst source and other
contributors within the field of view when determining $f_{\rm a}$, in
contrast to \cite{worpel2013,worpel2015}.

We fit the burst spectra with a combination of the pre-burst model
(disk blackbody and power law), fixing its parameters to the pre-burst
values and multiplying it with a free constant equal to $f_{\rm a}$,
and a Planck function for the burst radiation with two free parameters
(normalization and temperature). The complete model was multiplied
with the absorption model for a cold interstellar medium by
\cite{mor83} and fixing the equivalent hydrogen atom number column
density $N_{\rm H}$ per source to the value tabled in
\cite{worpel2013}.

A problem with the $f_{\rm a}$ model is \citep[see also][]{worpel2015}
that the shapes of the blackbody and non-burst spectra are rather
similar within the PCA's 3--20 keV calibrated bandpass, even more so
if the statistical quality is not high such as during short
exposures. Consequently, there is a coupling between the
normalizations of the two components that introduces additional
uncertainty in the luminosity of the blackbody component \citep[up to
  tens of percents according to][]{worpel2015}. This decreases the
diagnostic power of the light-curve analysis and increases the
uncertainties of the parameters.

All spectral fits were applied using XSPEC version version 12.9.1
\citep{arn96} in the {\em \textup{calibrated}} PCA bandpass of 3--20
keV and included a 0.5\% systematic uncertainty
\citep{sha12}.\typeout{NUMBERS} A total of more than 35,600 burst
spectra were generated. For two-thirds of these spectra, the source
was significantly detected in the sense that the flux was at least
three times higher than its uncertainty. For these, 4\% of the
spectral fits had $\chi^2_\nu>2$ with an average number of degrees of
freedom of $\langle \nu \rangle=18.7$. This percentage is 20 times
larger than expected for purely statistical
fluctuations. Figure~\ref{figexamplesp} shows the time-resolved
spectroscopy of an example burst from GS~1826-24, detected on
MJD~51811.74968 (September 24, 2000, at 17:59:32 UTC) with PCUs 0, 2,
and 3.

From the fitted blackbody parameters, we calculated the bolometric
flux. The uncertainty in the bolometric flux was determined by Monte
Carlo sampling of the temperature-normalization plane, calculating the
bolometric flux for each sample and determining the range of the
central 68\% values. We note that the bolometric correction,
applicable from the 3--20 keV band, is between 1.15 and 1.65 for
blackbody temperatures of 3 and 1 keV, respectively.

\subsection{Caveats}
\label{sec:caveats}

Our approach to the spectral modeling is, except for the free $f_{\rm
  a}$, traditional and effective. Nevertheless, there are some
caveats.  Foremost, the effective temperature and normalization
derived from the blackbody fit are mere proxies for the true effective
temperature and emission size. Neutron star surfaces are not
blackbodies, but atmospheres that are (nearly) completely ionized and
where particularly the hot electrons Comptonize the radiation from
below. Furthermore, the atmosphere may contain an increased level of
metals that influence the blackbody spectrum through free-free and
bound-free electron-ion interactions. Radiation transfer calculations
of model atmospheres \citep{lon86,mad04,maj05,sul11,sul12,nattila2015}
show that the continuum spectrum may be described by a 'diluted'
blackbody with a normalization correction factor $w$ (usually $w<1$,
therefore 'diluted') and a color-correction factor $f_{\rm c}$
(usually $f_{\rm c}>1$) to calculate the observed normalization and
'color' temperature from the true Planck normalization and effective
temperature. If the bolometric flux were unaffected, $f_{\rm c}^4w$
would be 1. \cite{nattila2015} calculated this number for a variety of
metallicities from 1 to 40 times solar plus a pure iron atmosphere,
and for gravitational accelerations $g$ between log$g$=14.0 and 14.6
for luminosities $L$ between 0.001 and 1 times the Eddington limit
$L_{\rm Edd}$. For $Z=Z_\odot$, we studied $f_{\rm c}^4w$ versus $L$
and find for $L>0.1L_{\rm Edd}$ that the trend is consistent with a
power-law function with an index of -0.03. We consider this a
systematic uncertainty in our cooling power-law decay indices that is
due to the blackbody model. Our data also cover the range $L<0.1L_{\rm
  Edd}$, but we refrained from considering this range because the
models show a strong decrease in observed luminosity (and therefore a
steepening of the decay), while the data generally do
not. \cite{nattila2015} attributed this steepening to an opacity that
is due to free-free interactions dominating the opacity due to
electron scattering.

While leaving $f_{\rm a}$ free improves spectral fits considerably, it
is expected that not only the normalization but also the shape of the
spectrum of the accretion-induced radiation should change during a
burst.  If the accretion disk corona is irradiated by the burst
photons, its temperature may adjust to the typical temperature of
these photons and its spectral shape will change.  If the
Poynting-Robertson effect is strong, the changing accretion rate is
expected to result in variable spectral shapes. We ignored these
possibilities because the current spectral model is already rather
satisfactory \citep[even in data in a wide
  bandpass;][]{zand2013}. This implies that it is difficult to
separate the accretion- and burst-induced spectra from each other,
given the limited bandpass and the similar shapes.

We assumed that the burst emission is isotropic. If any anisotropy
exists and changes during the burst, this may affect the light
curves. However, the few percent anisotropies suggested from burst
oscillation measurements \citep[e.g.,][]{wat12} are largely averaged
out during 1s time bins by the rotation of the neutron star, which
exceeds $10$~Hz rotation.

We neglected any absorption features in the burst spectra that might
change in amplitude because the ionization degree of the photosphere
changes as a result of cooling. Such features have been detected in a
few particularly powerful bursts
\citep[e.g.,][]{stro02a,jvp90,intzand2010,keek2014,kajava2017b}. However,
they make up less than $\sim0.1$\% of the PCA burst population.

We assumed that the spin of the neutron star has no noticeable effect
on the luminosity, either through rotational broadening of spectra by
special and general relativistic effects, from an oblateness of the
neutron star surface or because burning is confined to changing areas
on the neutron star. Bursting neutron star spins are limited to at
most 620 Hz, yielding maximum velocities at the equator of 0.1$c$ and
an oblateness that is predicted to be a few percent at most
\citep[e.g.,][]{baubock2012}. The implications for the bolometric flux
are predicted to be limited to a few percent at most and are constant
\citep[e.g.,][]{baubock2015}.

\section{Bolometric light-curve analysis}
\label{secanalysis}

The goal of our analysis is to model the decay in the bolometric flux
of type I X-ray bursts. This bolometric light curve results from the
time-resolved spectral modeling discussed in the previous section. In
this section we first introduce the approach taken for the light-curve
modeling and then discuss a high-quality test case.

\subsection{Approach and caveats}
\label{sec:approach}

\begin{figure}[t]
\begin{center}
\includegraphics[width=\columnwidth,angle=0,trim=1.5cm 2cm 1.5cm
  0cm]{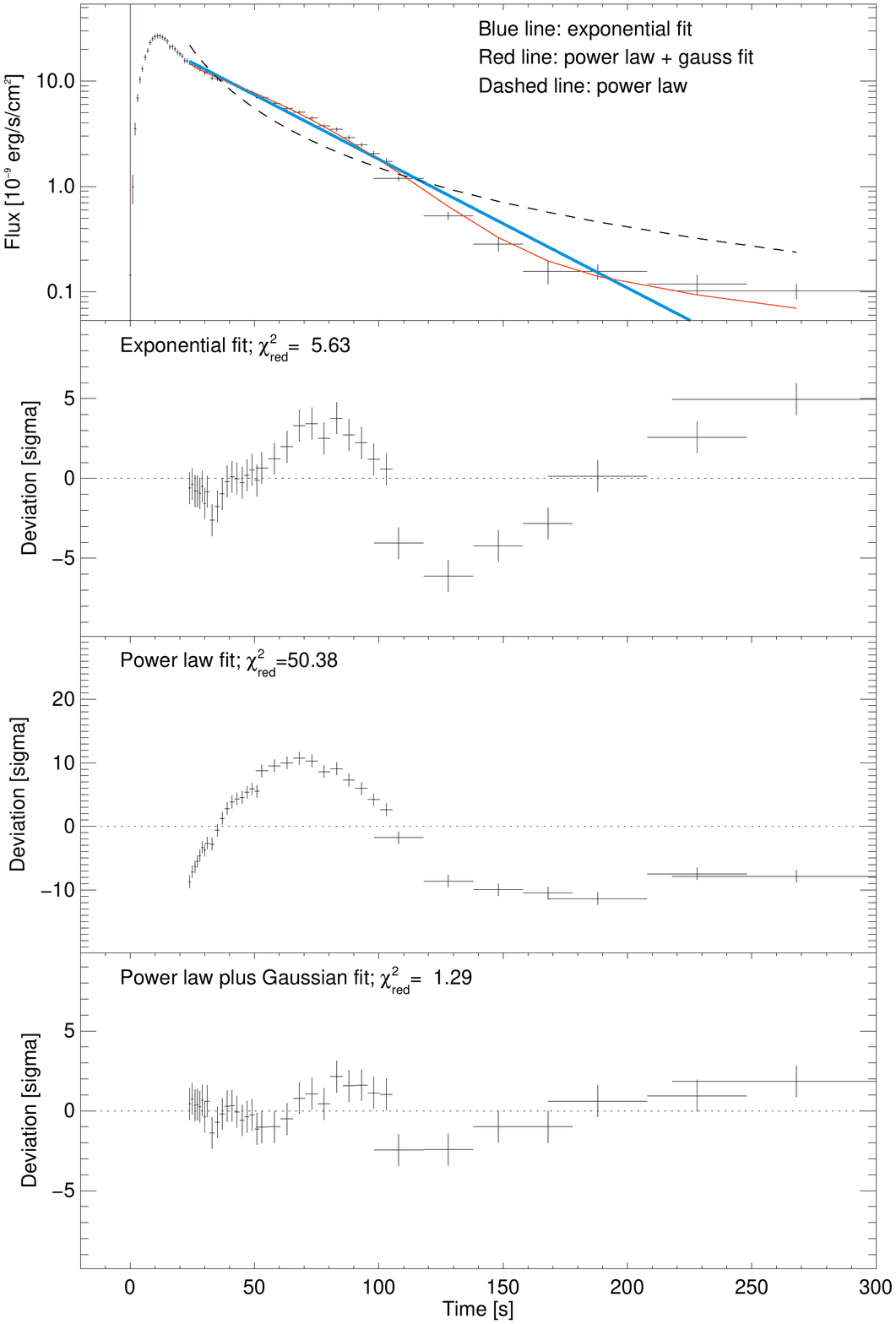}
\end{center}
\caption{Application of modeling on the average light curve of 29 RXTE bursts
from GS 1826-24 as determined in \cite{zan09}.\label{figtestcase}}
\end{figure}

The approach that we take to model the bolometric light curves is
empirical. Instead of attempting to explain the light curves from
first physics principles, which is non-trivial because it involves
many different and mutually dependent physical processes whose
magnitudes are sometimes ill-constrained
\citep[e.g.,][]{woosley2004,cyburt2016}, our approach is to apply the
simplest mathematical model that is consistent with the data, but has
a basic relationship to the physics in terms of the number of
components and first-order mathematical forms. Our principal aim is to
distinguish the effects of neutron star cooling from those of the rp
process. Therefore, our mathematical description consists of two
components with different mathematical forms.

Our analysis of the bolometric light curve of each burst consists of
the following steps: \newcounter{ll2}
\begin{list}{\arabic{ll2}.}{\leftmargin=0.4cm \parskip=0cm \partopsep=0cm \itemsep=0cm \parsep=0cm \topsep=0.cm \usecounter{ll2}}
\item Double check the start of the burst visually, redetermine start
  times if needed (by at most a few seconds), redefine all times with
  respect to this start time.
\item Determine the burst peak flux.
\item Determine the first data point of the decay to be included in
  fitting the model. This is defined as the last point that drops
  below 55\% of the peak flux (note that the flux may fluctuate and
  cross the 55\% mark multiple times). The 55\% mark was evaluated to
  be the most practical in terms of separation to possible dynamic
  effects of the emission region (photospheric expansion). Moreover,
  certain bursts, for instance most from XTE J1814-335 \citep{smsz03},
  clearly have two components: a fast initial spike, and a gradual
  decrease. The time of this 55\% mark is referred to by $t_0$. In 10\%
  of the cases, manual adjustments had to be made to $t_0$ because the
  light-curve behavior differed near the peak, see
  Appendix~\ref{appc}.
\item \label{item3p} Fit the decay data with three models: 
\begin{enumerate}
\item[a.] The traditional two-parameter exponential function
\begin{eqnarray}
F_1(t)=F_1(t_0) \; {\rm e}^{-(t-t_0)/\tau}, 
\label{eqn1}
\end{eqnarray}
 with normalization $F_1(t_0)$ and exponential decay time $\tau$.
\item[b.] The two-parameter power law
\begin{eqnarray}
F_2(t)=F_2(t_0)  \left( \frac{t}{t_0} \right)^{-\alpha_2}
\label{eqn2}
\end{eqnarray}
 with normalization $F_2(t_0)$ and decay index $\alpha_2$.
\item[c.] The four-parameter combination of a power law and a
  one-sided Gaussian function
\begin{eqnarray}
F_3(t) = F_3(t_0) \left( \frac{t}{t_0} \right)^{-\alpha_3} + \frac{G}{\sqrt{2\pi}s} \;
{\rm e}^{-\frac{t^2}{2s^2}}
\label{eqn3}
,\end{eqnarray} with $G$ the normalization of the Gaussian function
and $s$ its standard deviation.  The Gaussian function centroid is
fixed to the start of the burst $t=0$, assuming that the temperature
and the hydrogen abundance in the burning layer, and therefore the rp
rate, are at a maximum at that time.
\end{enumerate}
\item Calculate the fluence by adding the integration of the fitted
  decay model, from the fit start time to infinity, to the fluences of
  all the data points before this until the start of the burst. For
  the third model, we separately calculate the fraction $f$ of the
  Gaussian fluence ($G/2$) to the total fluence.
\end{list}

\noindent
The functional forms employed in item \ref{item3p} are motivated by
the tradition in burst analyses to model decays with an exponential
function, the theoretical expectation that the cooling process follows
a power law \citep{cumming2004,zand2014a}, and our goal to model any
residuals that look like bumps by the simplest function possible,
which is a two-parameter Gaussian function. The choice for a Gaussian
function is completely empirical. When we tested this function with
theoretical models for the rp process, we found some justification;
see Appendix~\ref{appa}. It provides a tool to quantify the complex rp
process in a straightforward and numerically robust manner that is
more or less consistent with the data. The Gaussian amplitude
indicates the amount of energy liberated by the rp process, and the
Gaussian width its timescale and, thus, nuclear chain length (see
below).

We note that we have experimented with several alternatives to the
one-sided Gaussian: 1) a Gaussian with a free centroid. This provides
somewhat better fits, but the parameters are much more difficult to
constrain since this Gaussian has a tendency to fit to any curvature
in the observed decay, for instance, the curvatures found later in the
burst when variations in the persistent flux may become important. 2)
A broken power-law function as a first approximation to a continuously
varying power law \citep[e.g.,][]{zand2014a}. This always fits worse.
3) The simplified rp model explained in Appendix~\ref{appa1}, with the
parameters amplitude and rp endpoint. We found that this model does
not fit the data well because the model timescales are insufficiently
long with respect to the data, see also Fig.~\ref{figrp}.

The various decay functions (Eqs.~\ref{eqn1}, \ref{eqn2} and
\ref{eqn3}) in relation to one particular burst are illustrated in
Fig.~\ref{figexample} (this is for the same burst whose time-resolved
spectroscopy is shown in Fig.~\ref{figexamplesp}). The exponential
decay is unsatisfactory. The power law even more so, showing a broad
excess centered at about 50 s that can be well modeled by including
half a Gaussian centered at burst onset. In general, it takes
accuracies of a few percent of the peak flux and a dynamic range in
excess of a factor of 20 to distinguish the Gaussian
component. Therefore it can only be detected with instrumentation with
the highest throughput, like the PCA, and not, for instance, with WFC
or JEM-X.

We note that best-fit parameters were determined through the
Levenberg-Marquardt method \citep[e.g.,][]{bevington}, with the
distinction that $G$ and $s$ were forced to be positive by employing
as fitting parameters $G_{\rm f}$ and $s_{\rm f}$ so that
$G=\sqrt{G_{\rm f}^2+1}-1$ and $s=\sqrt{s_{\rm f}^2+1}-1$.

In a power law, the applied start time influences the inferred value
for $\alpha$ \citep[see Fig.~3 of][]{zand2014a}. Roughly speaking, an
uncertainty of 1 s in the start time translates into an uncertainty of
0.1 in $\alpha$. We consider the employed start times to be generally
more accurate than 1 s. In 0.4\% of the cases we manually adjusted the
burst start time to accommodate a better fit (see Appendix
\ref{appc}).

\subsection{High-quality test case}
\label{sec:testcase}

We verified our light-curve modeling on a high-quality test case from
the RXTE data archive. GS 1826-24 is a well-documented X-ray burster
with highly reproducible bursts and only very slow changes of
non-burst emission \citep{ube99,gal04,heg07a}. \cite{zan09} employed
these characteristics to generate a high-quality burst light-curve by
averaging PCA/PCU2 data of 29 bursts, to study the tail of these
bursts on much longer timescales. We employed our analysis on this
very same light curve. The result is shown in
Fig.~\ref{figtestcase}. We find that our model 3 is consistent with
these data, although there are some small systematic deviations in the
data (see bottom panel of figure).

\section{Results}
\label{results}

\begin{figure}[t]
\begin{center}
\includegraphics[width=0.93\columnwidth,angle=0,trim=1.5cm 7.5cm 1.5cm
  6cm]{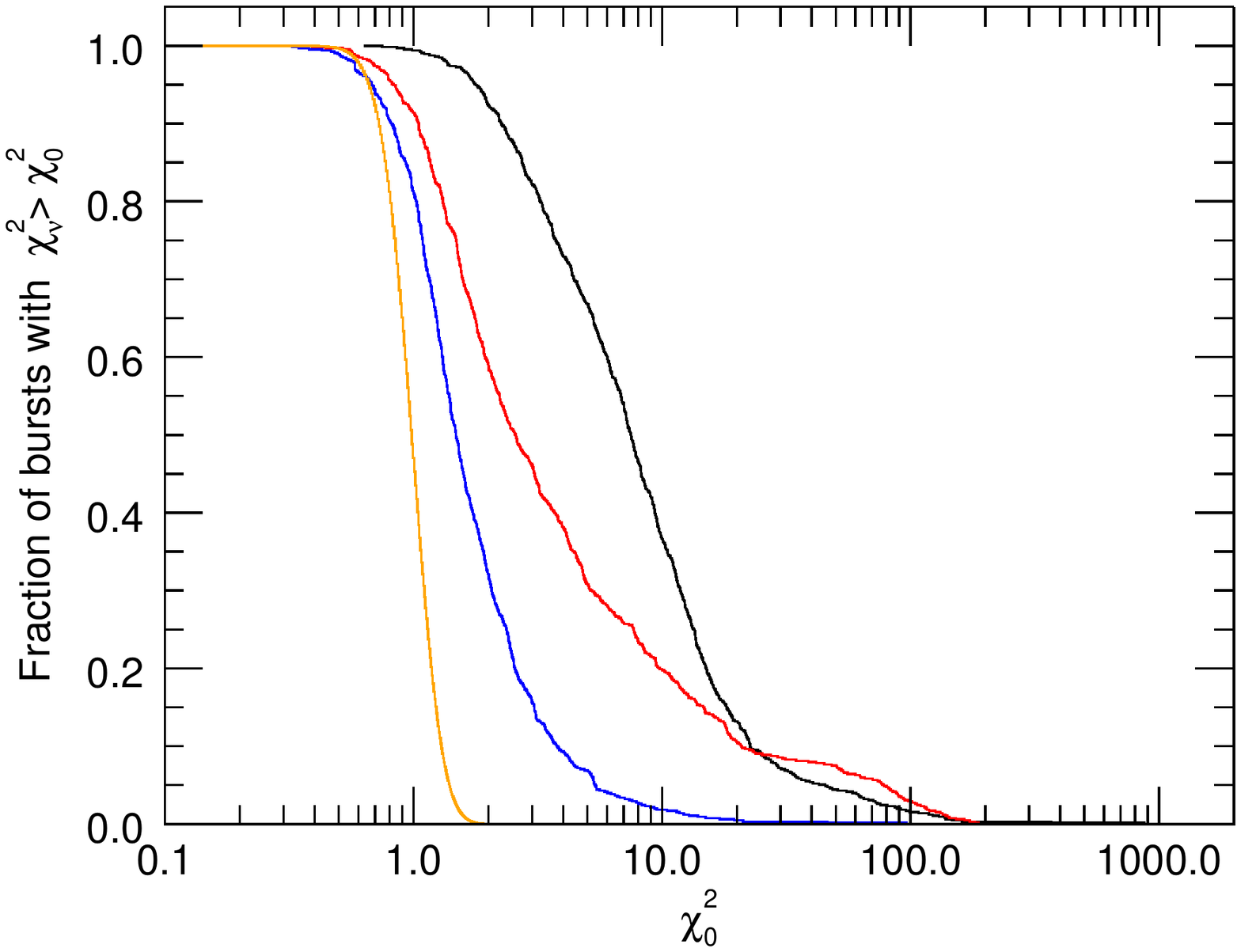}
\end{center}
\caption{Cumulative distribution for 1254 bursts of $\chi^2_\nu$ for
  fits with Eq.~\ref{eqn1} (black), \ref{eqn2} (red), and \ref{eqn3}
  (blue). The yellow curve shows the theoretical $\chi^2_\nu$
  distribution for 40 degrees of freedom (which is the
  average).\label{fig1}}
\end{figure}

Figure~\ref{fig1} shows the cumulative distributions of the reduced
chi-squared $\chi^2_\nu$ for the three models.  We find that the whole
burst sample is best fit with the power-law plus Gaussian model and
worst fit with the traditional single-exponential function. The
distribution for the third model is also closest to the theoretically
expected distribution (yellow curve, this is the $\chi^2$ distribution
for Gaussian statistics assuming 40 degrees of freedom, which is the
average value).

While the power-law plus Gaussian model is generally the best, it is
formally not consistent with the data, given the discrepancy in the
cumulative $\chi^2_\nu$ distribution.  When fixing $f_{\rm a}$ to 1,
this discrepancy worsens. The smaller discrepancy for a free $f_{\rm
  a}$ may be explained by the aforementioned degeneracy between the
persistent and burst spectrum in the 3--20 keV band. The inconsistency
of the best-fit model may be partly due to a short-term variability in
the spectral shape of the persistent emission. We account for the
implied uncertainty by multiplying all associated parameter
uncertainties with $\sqrt{\chi^2_\nu}$ for the power law or the power
law plus Gaussian if $\chi^2_\nu>1$.

We review further results through histograms and diagrams. We include
only bursts for which the error in the power-law index is smaller than
0.2, the power-law index itself is between 1.0 and 2.5, the error in
the Gaussian width $s$ is smaller than 50 s, and the error in the
fluence fraction is smaller than 0.1. A final selection was applied by
requiring that the burst peak flux is at least 15 times higher than
the pre-burst flux. This avoids the worst case of degeneracy between
the accretion and burst spectrum, which would affect the calculation
of the bolometric burst flux, but remains sensitive enough to detect
the Gaussian component. All these filters decrease the sample by
approximately half to 501. \typeout{NUMBERS}

\begin{figure}[t]
\begin{center}
\includegraphics[width=0.93\columnwidth,angle=0,trim=1.5cm 7.5cm 1.5cm 6cm]{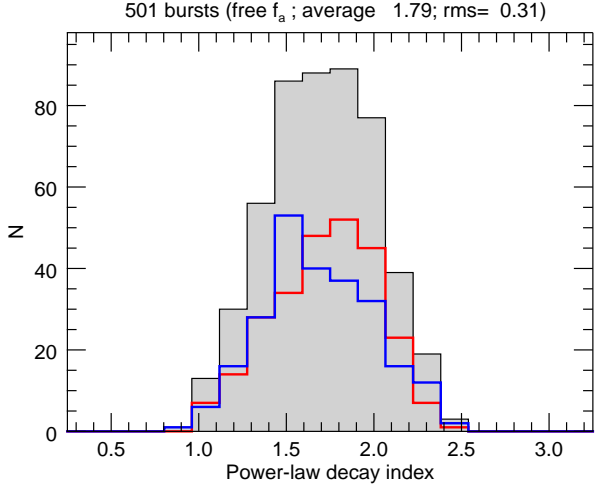}
\end{center}
\caption{Histogram of power-law decay index $\alpha$ (gray shaded)
  over all acceptably fitting bursts, as found from fitting with
  Eqs.~\ref{eqn2} ($\alpha$=$\alpha_2$) or \ref{eqn3}
  ($\alpha$=$\alpha_3$) when Eq.~\ref{eqn2} did not yield a
  satisfactory fit (i.e., if $\chi^2_\nu>2$), Eq.~\ref{eqn3} yielded a
  better fit , and $\chi^2_\nu$ for the fit was less than 10. The red
  histogram is for bursts without a noticeable Gaussian component (259
  bursts), and the blue histogram shows bursts with a noticeable
  Gaussian component (242 bursts)\typeout{NUMBERS}. \label{fig3}}
\end{figure}

Figure~\ref{fig3} shows the histogram of the fitted power-law decay
index. The index was taken from the fit with the power law
(Eq.~\ref{eqn2}) if $\chi^2_\nu<2$, or from that with the power law
plus Gaussian (Eq. ~\ref{eqn3}) if its $\chi^2_\nu$ was smaller than
that for the fit with the power law. Eighty percent of the decay
indices range from roughly 1.3 to 2.1. The average decay index is 1.79
and the {\rm rms} is 0.31.

\begin{figure}[t]
\begin{center}
\includegraphics[width=0.93\columnwidth,angle=0,trim=1.5cm 7.5cm 1.5cm 6cm]{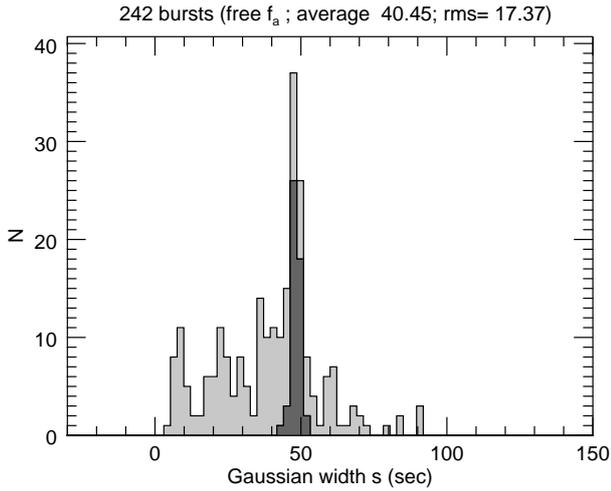}
\end{center}
\caption{Histogram of Gaussian width $s$, as found from fitting with
  Eq.~\ref{eqn3}. The peak at 48 s is due to bursts from GS 1826-24,
  as indicated by the dark gray shaded part of the histogram. Note the
  cutoff at 60-70 s.\label{fig4}}
\end{figure}

\begin{figure}[t]
\begin{center}
\includegraphics[width=0.93\columnwidth,angle=0,trim=1.5cm 7.5cm 1.5cm 6cm]{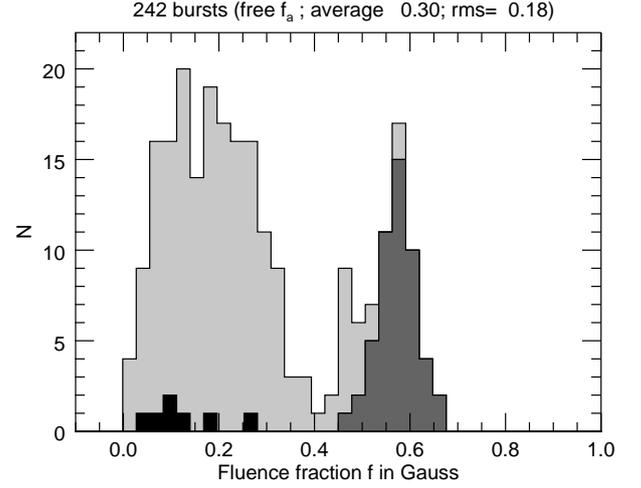}
\end{center}
\caption{Histogram of the fraction $f$ of the fluence contained in the
  Gaussian component. The peak at 0.6 is due to bursts from GS
  1826-24, as indicated by the dark gray shaded part of the
  histogram. The black histogram blocks are due to bursts from UCXBs
  at $f<0.3$. Note the cutoff beyond the rightmost peak.\label{fig6}}
\end{figure}

\begin{figure}[t]
\begin{center}
\includegraphics[width=0.93\columnwidth,angle=0,trim=1.5cm 7.5cm 1.5cm 4.5cm]{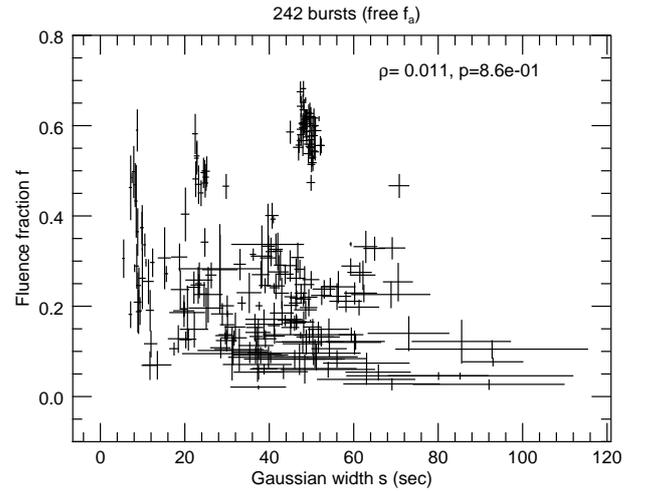}
\end{center}
\caption{Diagram of $f$ against $s$ for bursts that fit
  Eq.~\ref{eqn3} best. Indicated in the top right corner are the Spearman
  rank-order correlation coefficient $\rho$ and the chance probability
  that $\rho$ is exceeded.\label{fig52}}
\end{figure}

\begin{figure}[t]
\begin{center}
\includegraphics[width=0.93\columnwidth,angle=0,trim=1.5cm 7.5cm 1.5cm 4.5cm]{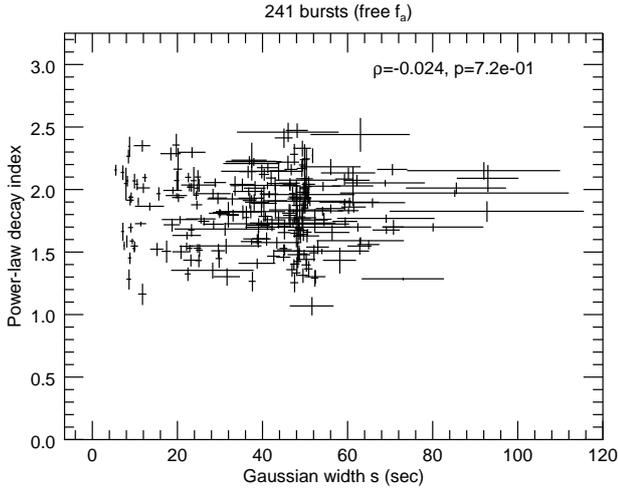}
\end{center}
\caption{Diagram of $\alpha_3$ against $s$ for bursts that fit best
  Eq.~\ref{eqn3} best.\label{fig5}}
\end{figure}

\begin{figure}[!t] % from !t to t for referee version
\begin{center}
\includegraphics[width=0.75\columnwidth,angle=0,trim=1.5cm 6.8cm 1.5cm 6cm]{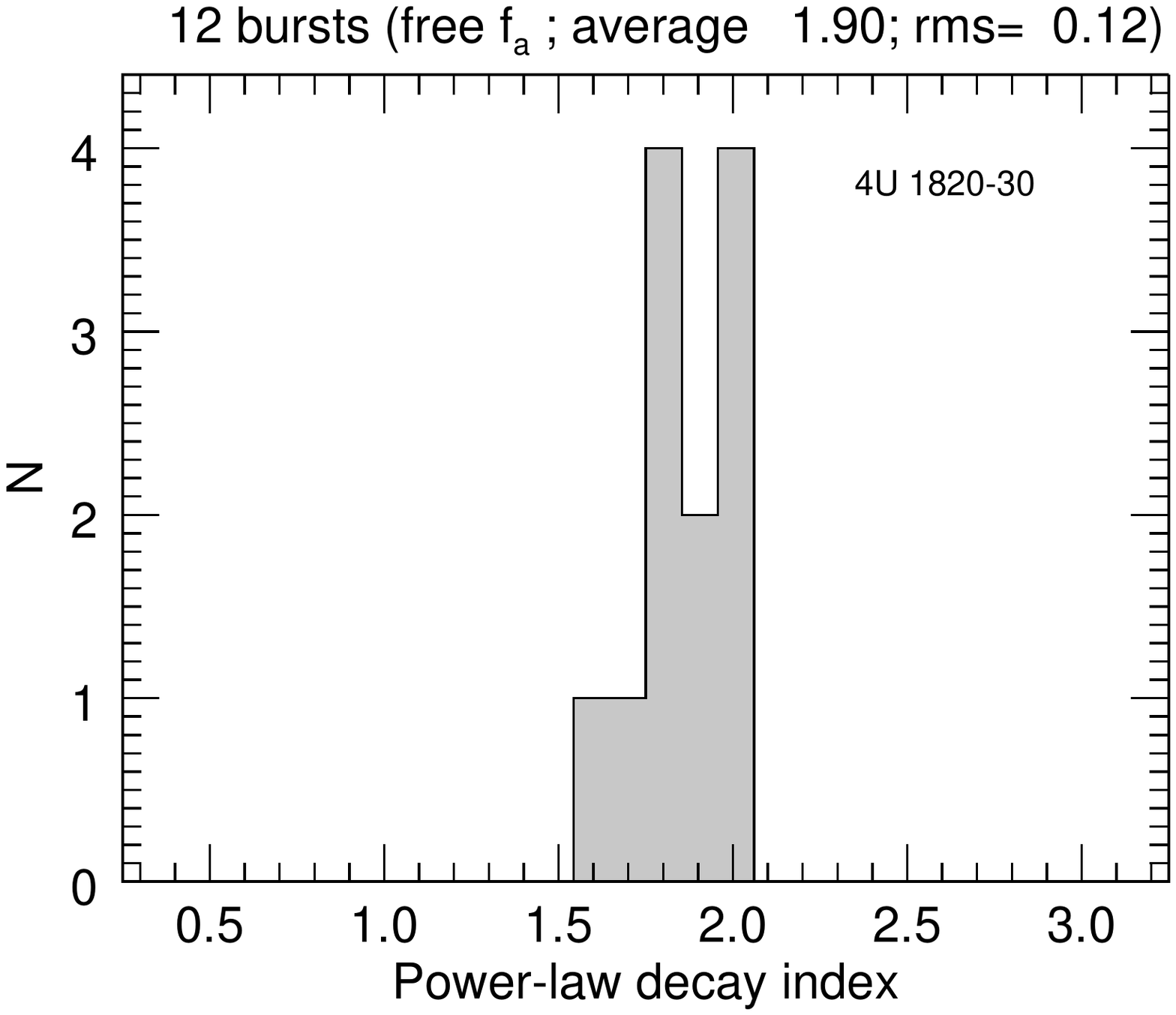}
\includegraphics[width=0.75\columnwidth,angle=0,trim=1.5cm 7.8cm 1.5cm 6cm]{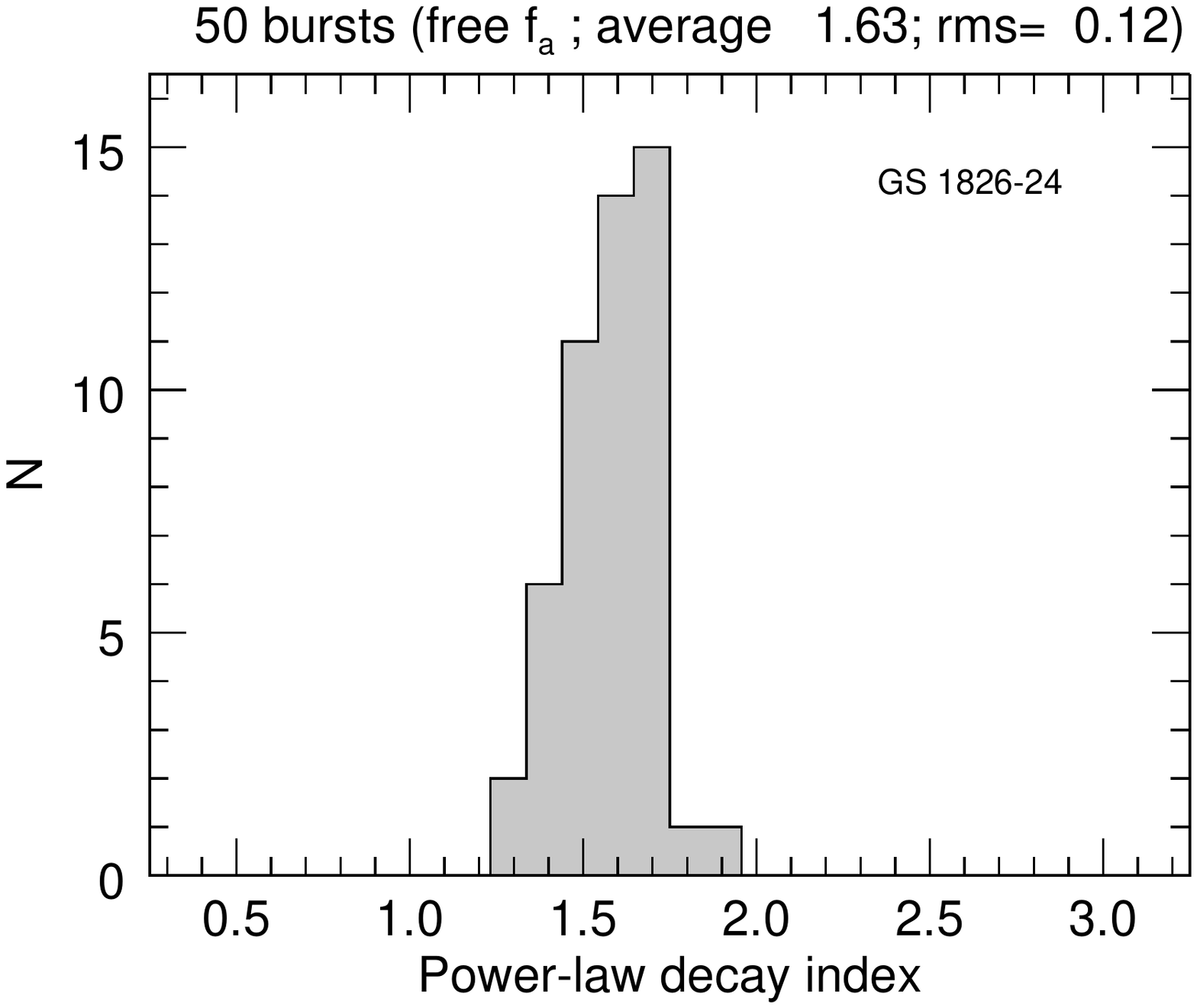}
\includegraphics[width=0.75\columnwidth,angle=0,trim=1.5cm 7.8cm 1.5cm 5cm]{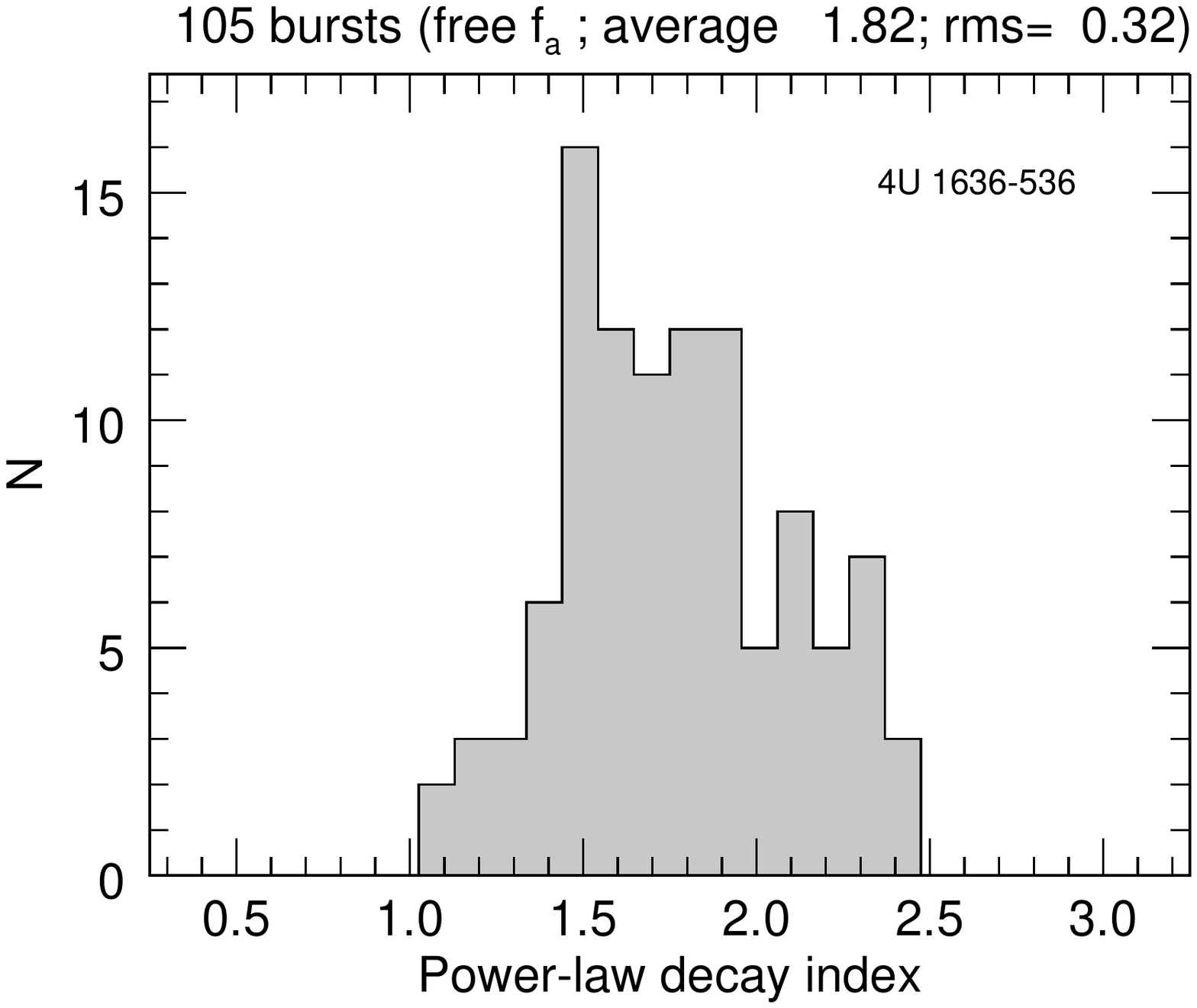}
\includegraphics[width=0.75\columnwidth,angle=0,trim=1.5cm 7.8cm 1.5cm 5cm]{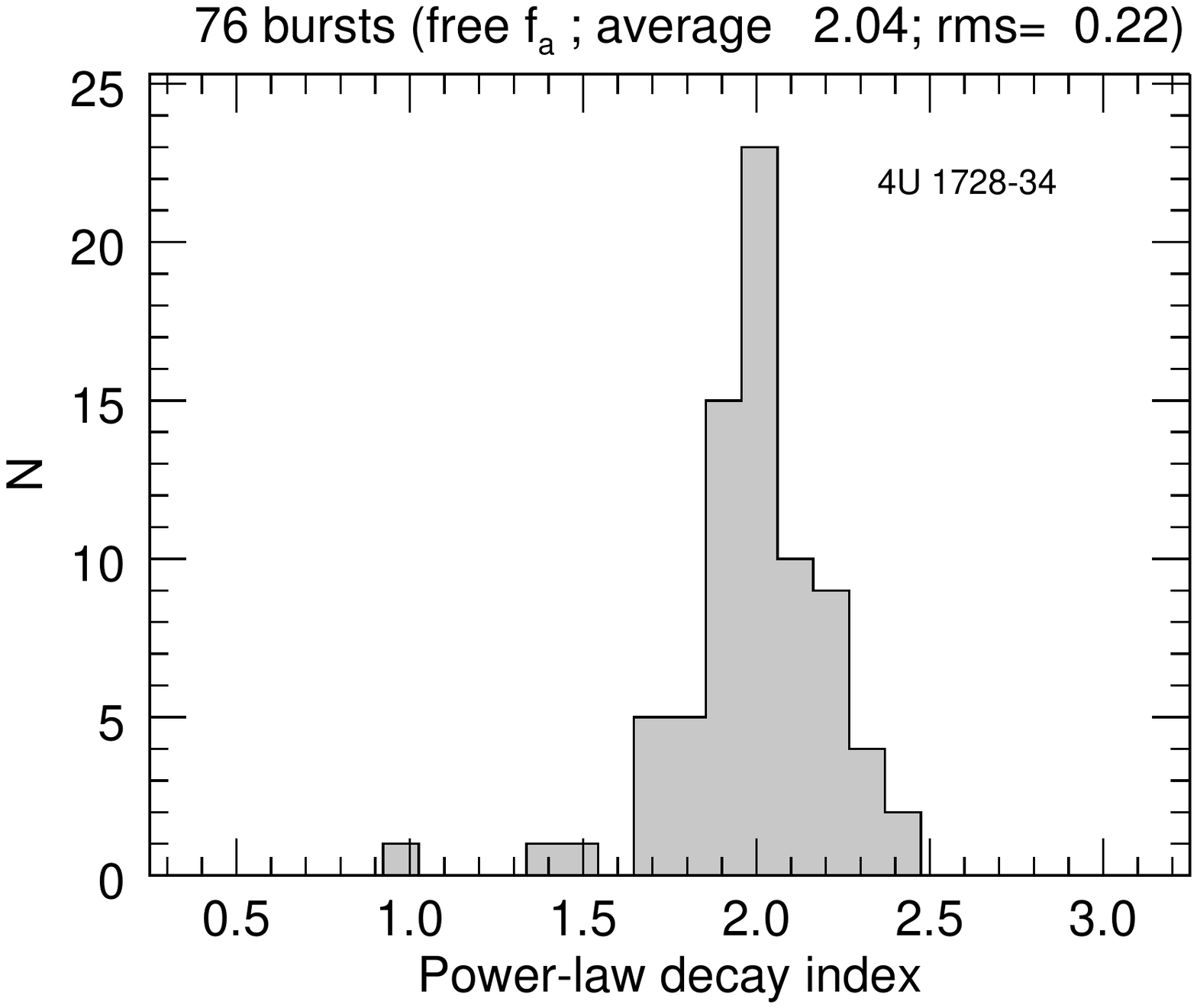}
\end{center}
\caption{Histograms of $\alpha$ for four bursters: two with small
  variations in persistent flux (4U 1820-30 and GS 1826-24), and two
  with large variations (4U 1636-536 and 4U 1728-34). 4U 1820-30 has a
  negligible hydrogen abundance (lower than 10\%) in the accreted
  matter; this is possibly also true for 4U 1728-34. \label{figpowc}}
\end{figure}

Two additional distributions are shown in Fig.~\ref{fig3}. The red
histogram shows the distribution of the bursts that lack a Gaussian
component. There is no qualitative difference with the histogram over
all bursts. The average of this distribution is 1.80 and the rms is
0.30. The blue diagram is for all bursts with a significant Gaussian
component, with an average of 1.77 and an rms of 0.32. The
Kolmogorov-Smirnov test between the data for the red and blue
histograms is 0.128; \typeout{NUMBERS} the probability that the value
is this high or higher by chance is 3\%. The two-sample
Anderson-Darling test \citep{engmann2011,pettitt1976} between both
histograms is $1.857$ with a chance probability of about 10\%. We
consider both these tests as insufficient evidence for a difference in
the two histograms.

\begin{figure}[t!] % from !t to t for referee version
\begin{center}
\includegraphics[width=\columnwidth,angle=0,trim=1.5cm 7.5cm 1.5cm 6cm]{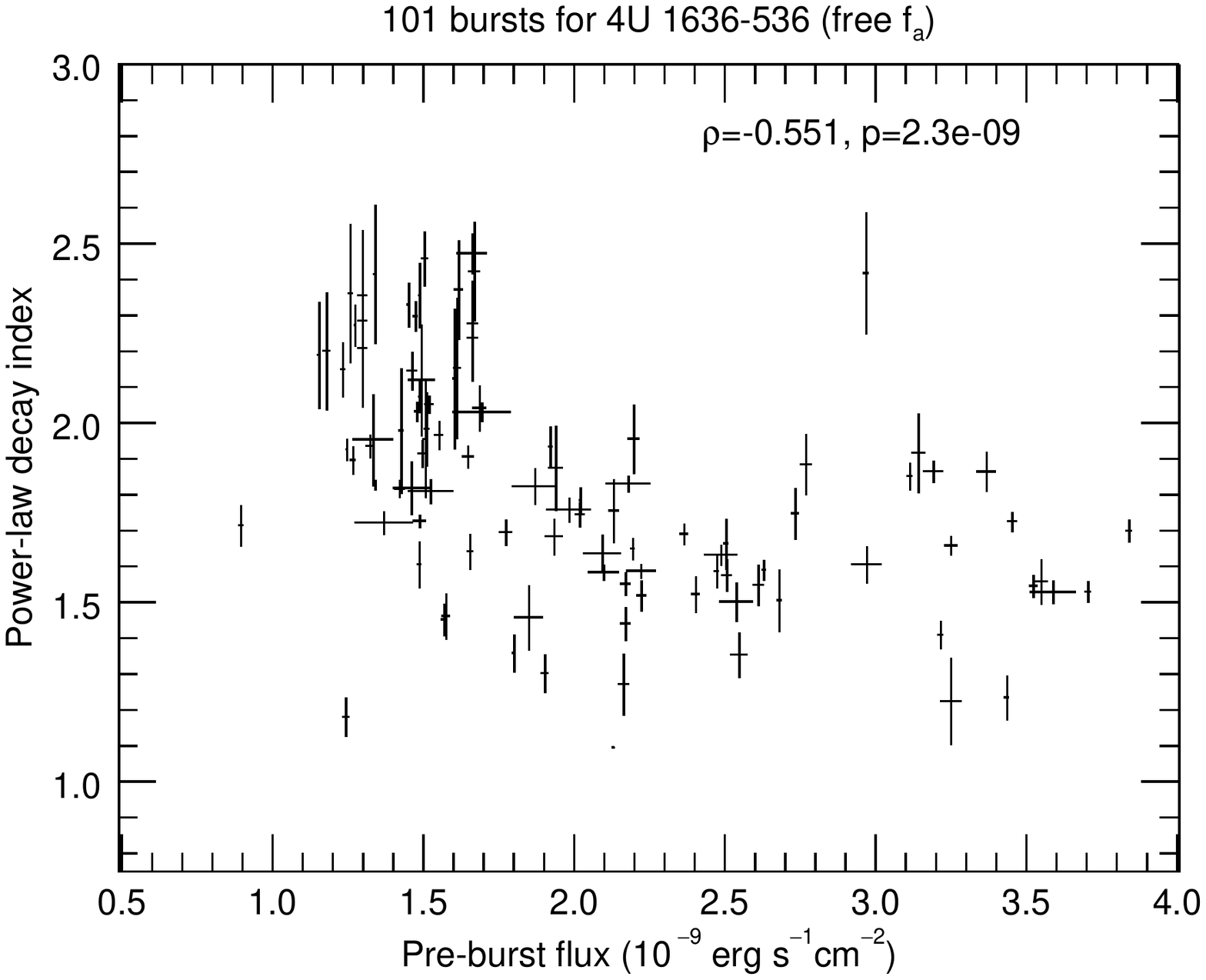}
\end{center}
\caption{Diagram for 4U 1636-536 of $\alpha$ against the pre-burst
  3--20 keV flux (in units of 10$^{-9}$~\ecs). \label{fig11}}
\end{figure}

\begin{figure}[!t] % from !t to t for referee version
\begin{center}
\includegraphics[width=\columnwidth,angle=0,trim=1.5cm 7.5cm 1.5cm 6cm]{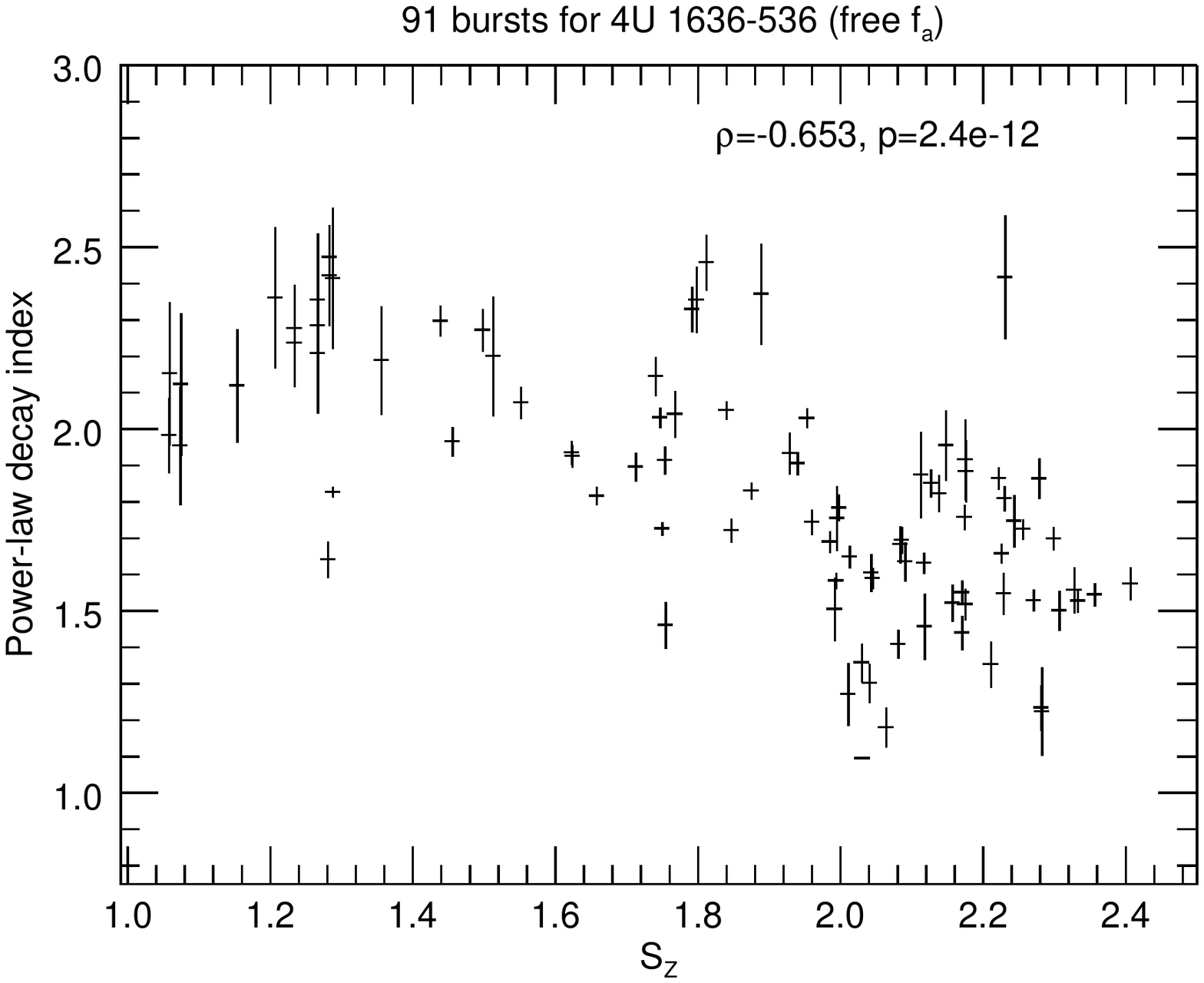}
\end{center}
\caption{Diagram for 4U 1636-536 of $\alpha$ against $S_{\rm z}$. The
  error margins on $S_{\rm z}$ are fiducial.\label{figzp}}
\end{figure}

We also separately studied the distribution of power-law indices for
all (candidate) UCXBs 4U 0513-40, 2S 0918-549, 4U 1820-30, 4U 1916-05,
4U 1728-34, and 4U 2129+11. We found 108 \typeout{NUMBERS} bursts for
these. One hundred and one are best fit with a single power law. The
decay index ranges for 80\% between 1.4 and 2.1 with an average of
1.94 and an rms of 0.27. These are similar values as for the complete
sample, maybe somewhat narrower and steeper.

\begin{figure}[!t] % from !t to t for referee version
\begin{center}
\includegraphics[width=\columnwidth,angle=0,trim=1.5cm 7.5cm 1.5cm 6cm]{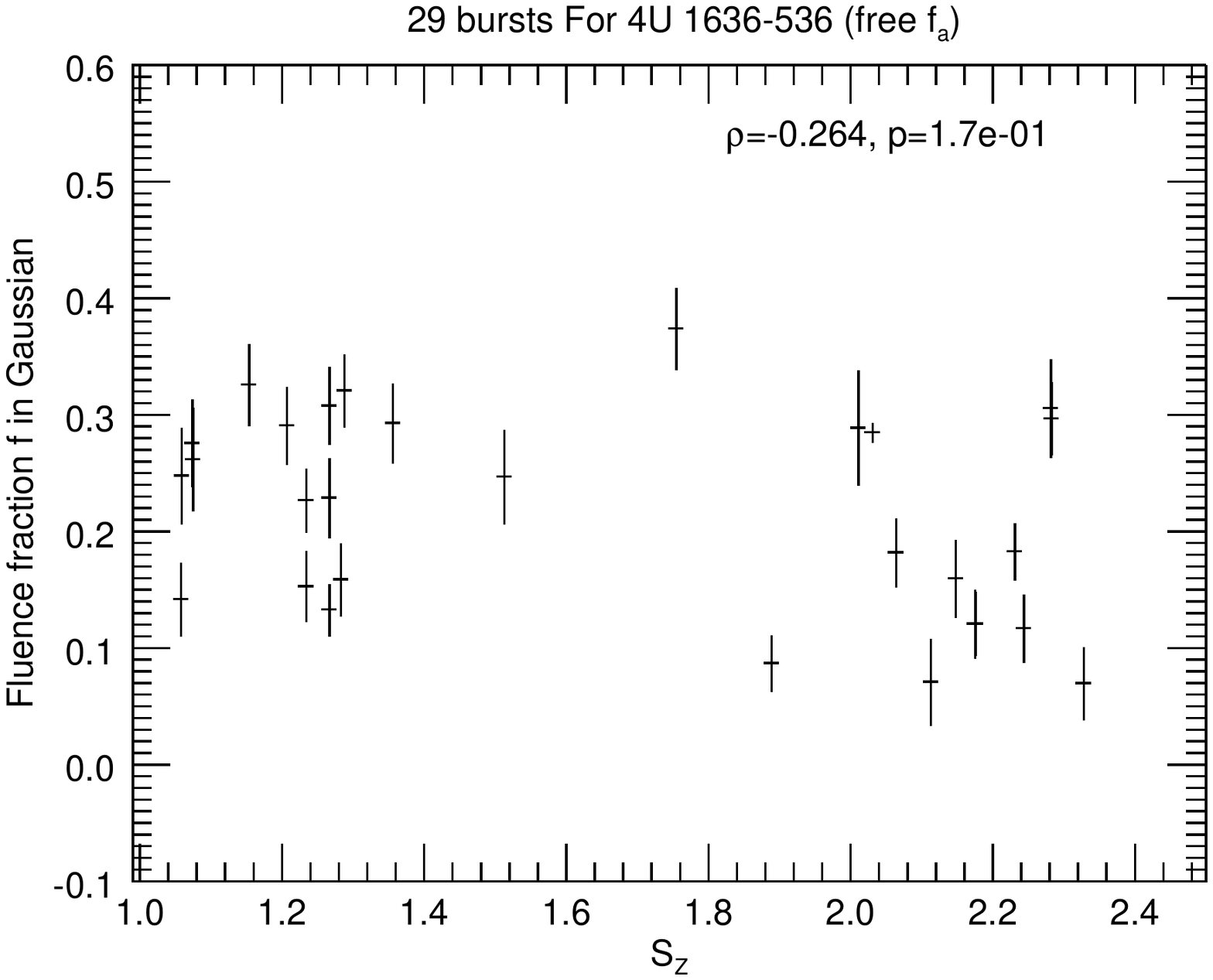}
\end{center}
\caption{Diagram for 4U 1636-536 of $f$ against the parameter $S_{\rm
    Z}$. The error margins on $S_{\rm z}$ are fiducial.\label{figz}}
\end{figure}

\begin{figure}[!t] % from !t to t for referee version
\begin{center}
\includegraphics[width=\columnwidth,angle=0,trim=1.5cm 7.5cm 1.5cm 6cm]{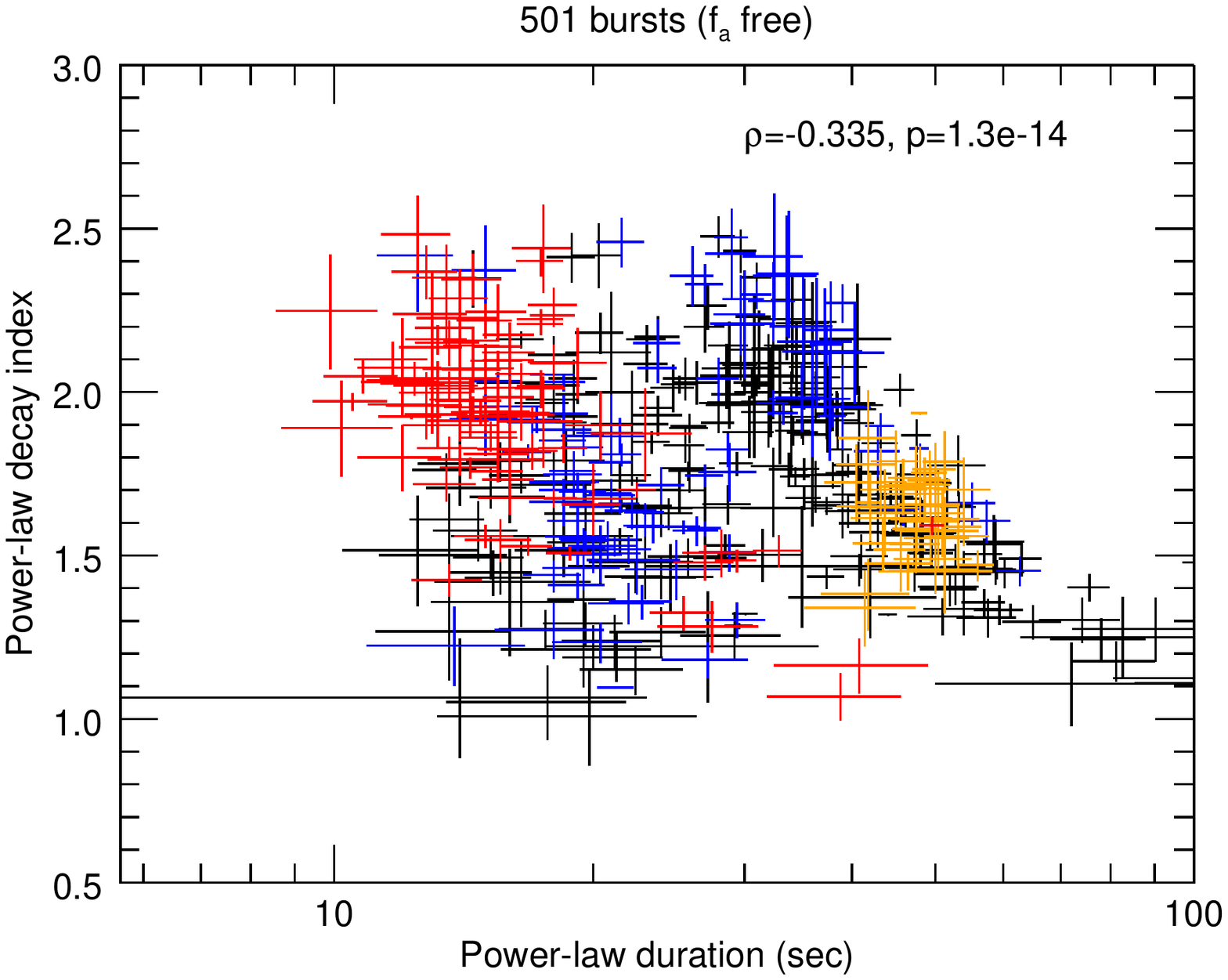}
\end{center}
\caption{Diagram of $\alpha$ against the duration when the power-law
  flux is above 5\% of the peak flux. The red data points refer to
  bursts from (hydrogen-deficient and, thus, rp-process deficient)
  UCXBs 4U 0513-40, 2S 0918-549, 4U 1728-34, 4U 1820-30, XB 1916-053,
  and 4U 2129+12. The blue data points refer to 4U 1636-536 and the
  orange points to GS 1826-24. This diagram shows the strongest trend in
  our data with two branches: the short (left) and long branch
  (right).\label{fig8}}
\end{figure}

Figure~\ref{fig4} shows the histogram of the fitted values for the
width $s$ of the Gaussian component for the bursts that better fit the
power law plus Gaussian. The distribution has one large peak at 48~s.
This is about half due to bursts from GS~1826-24. The distribution
cuts off above 60-70 s, well below the range covered by the data (300
s), with a few detections at large width that have large uncertainties
and are associated with bursts with a relatively small Gaussian
component ($f<0.1$, see Fig.~\ref{fig52}).

Figure~\ref{fig6} shows the distribution of the fraction $f$ of the
fluence contained in the Gaussian component. This shows that the
fluence of the Gaussian is mostly smaller than that of the power-law
component. This yields an interesting conclusion about the energy
liberated by the rp process, which we discuss in
Sect.~\ref{secdiscussion}. The distribution is bimodal with a broad
component that peaks at $f=0.2$ and decreases to almost zero at
$f=0.4$, and a narrow component centered at $f=0.6$. The latter
component is again foremost due to GS~1826-24. Both this histogram and
that of the width $s$ express the special case of GS 1826-24, which is
probably due to its peculiarly constant behavior \citep[see
  also][]{ube99,gal04,zan09,chenevez2016}. Seven out of 108 bursts
from UCXBs have a non-zero $f$ value, see Fig.~\ref{fig6}. Five are
from 4U 1728-34 and one from XB 1916-053 and 4U 0513-40. In all of
these cases, the fits with Gaussian components were only marginally
better than without. Figure~\ref{fig52} shows a diagram of $f$ versus
$s$. This shows that for each Gaussian width, there is a range of
fluence fractions contained in the Gaussian.

An interesting question is whether there is a correlation between
$\alpha$ and $s$ or $f$, because if there were, it would point to a
dependence of the cooling on the extent of rp processing. For the same
set of bursts, these two parameters are graphed against each other in
Fig.~\ref{fig5}. There is no obvious correlation with a Spearman
rank-order correlation coefficient of $\rho=-0.024$ and a chance
probability of $p=0.72$ for this value or larger. The same applies to
the dependence between $\alpha$ and $f$ ($\rho=-0.119$ and
$p=0.07$). This testifies to a bias-free evaluation of the power law
and Gaussian component.

In Fig.~\ref{figpowc} we show the distributions of the decay index for
four interesting sources with many bursts: GS 1826-24, 4U 1636-536, 4U
1820-30, and 4U 1728-34. Although 4U 1820-30 has considerably fewer
bursts than the others, we include it here because it is a confirmed
UCXB with a low hydrogen abundance in the accreted gas
\citep{cum03}. The distribution is narrow for GS 1826-24, consistent
with its constant behavior. The average index is $1.63$ and the rms is
0.12 \typeout{NUMBERS} (we note that the typical uncertainty in the
decay index is of similar magnitude). The distribution for 4U 1636-536
is three times broader and almost equal to that of all bursts. The
average index value is 1.82 \typeout{NUMBERS}. This source is known to
be very variable \citep[e.g.,][]{shih2011}. 4U 1820-30 has a narrow
distribution with an average similar to that of 4U 1636-536
(1.90\typeout{NUMBERS}). 4U 1728-34 has an intermediately broad
distribution with a steep average index of 2.04.

In Fig.~\ref{fig11} we show the diagram of the decay index of bursts
from 4U 1636-536 versus the 3--20 keV flux in the pre-burst spectrum
(in units of 10$^{-9}$~\ecs). There seems to be a trend between the
two parameters. Generally, steeper decay indices are found at low
pre-burst fluxes.  There is a stronger trend if we use $S_{\rm z}$
instead of pre-burst flux as ordinate, see Fig.~\ref{figzp}. $S_{\rm
  z}$ is a parameter inferred from the color-color diagram and is
interpreted as a better proxy for the mass accretion rate than flux
\citep[e.g.,][]{galloway2008}. It is not transferable from source to
source, however. There is also a small trend in the diagram of $f$
versus $S_{\rm z}$, see Fig.~\ref{figz}.  At low $S_{\rm z}$ values
(i.e., smaller than 0.1), there is a shortage of low $f<0.1$ values.

Hardly any trends are observed for the Gaussian fluence fraction $f$
(Fig.~\ref{figz}) and the Gaussian width $s$ versus pre-burst flux
(not shown).

In Fig.~\ref{fig8} we show the diagram of the decay index versus the
duration of the power-law decay, defined as the time that the
power-law flux remains above 5\% of the peak flux value following the
best-fit model for the decay. This diagram is not random; there is a
strong correlation between the two parameters. This is probably due to
a coupling between these parameters: a shallow decay automatically
results in a longer burst duration.  It is more interesting that there
appear to be two parallel branches. We tested whether they are related
to specific types of sources and found that bursts from 4U 1636-536
(the blue data points) spread over both branches, that bursts with a
significant Gaussian component also spread over both branches, but
that all 108 UCXB bursts except one are on the short
branch\typeout{NUMBERS}. It is also noticeable that all bursts from
the GS 1826-24 cluster are located on the long (right) branch. The
existence of the long branch is related to the fact that a large part
of the bursts remains near the peak for some time, even when the
Eddington limit is not reached, or has a long rise time. Clear
examples of this are bursts from the Rapid Burster
\citep{bag13,bag14}, GS 1826-24 (see Fig. \ref{figexamplesp}), and 4U
1636-536 \citep[e.g.,][]{zhang2009}.

Many histograms and diagrams show clustering. This may partly be
explained by the dominance in the number of bursts from the persistent
accretors 4U 1636-536, 4U 1728-34, and GS 1826-24 (see
Table~\ref{tab1}, column b).

\section{Discussion}
\label{secdiscussion}

With reasonable success, we have modeled the decay in bolometric flux
of a large sample of thermonuclear X-ray bursts by the combination of
a power law with a decay index that is constant over the burst, which
we regard as representative of cooling in the neutron star envelope,
and a one-sided Gaussian, representative of the rp process. One may
question whether these representations are valid since the cooling is
expected to be a power law with a varying decay index and the rp
process is expected to have more complicated light curves. However,
more appropriate functions for the cooling and rp process are
difficult, if not impossible, to constrain and distinguish with the
best data currently available. Since our simplified modeling does not
have these difficulties and is fairly successful, it is interesting to
report it and discuss implications, as far as possible.

Recently, \cite{kuuttila2017} also presented a study of X-ray burst
decay in a sample of 540 bursts detected with the PCA. Their focus was
on the bursts from 4U 1608-52, 4U 1636-536, 4U 1728-34, 4U 1820-30,
and GS 1826-24. This sample pertains to a subset of our sample of 1254
(cf. Table~\ref{tab1}). Their data treatment is different from ours,
which makes it interesting as a cross check against our
results. Basically, the treatment of \cite{kuuttila2017} follows the
$f_{\rm a}=1$ method for spectral modeling of the burst spectra,
employs the Planck function for the burst emission, and applies a
power law to the derived bolometric flux decays with a variable decay
index.  Complete burst profiles are also included, in contrast to our
restriction to fluxes below 55\% of the peak. \cite{kuuttila2017} made
no allowance for a second model component next to the power law, like
a Gaussian. They found that 1) for the (presumably) H-poor bursts from
4U 1820-30, $\alpha$ has a rather flat profile below about one-third
of the peak flux with a value of about 2.0 (cf. Fig.~\ref{figpowc});
2) for Eddington-limited bursts (i.e., bursts with photospheric radius
expansion or 'PRE') from 4U 1636-526 and 4U 1608-52, the same applies
(although at different $\alpha$ values); 3) there are strong $\alpha$
evolutions for all bursts from GS 1826-24 and non-PRE bursts from 4U
1636-536 and 4U 1608-52; 4) below one-third to one-fifth of the peak
flux, the $\alpha$ profile is flatter for bursts detected in the hard
state than in the soft state. Interestingly, \cite{kuuttila2017}
directly compared the $\alpha$ evolution with the theoretical
predictions from \cite{zand2014a} and found that they are generally
inconsistent for fluxes above one-third of the peak flux and
marginally consistent below this, with the exception of GS 1826-24,
which is strongly inconsistent throughout. In our opinion, both
studies (\citealt{kuuttila2017} and ours) show that a large portion of
the signal in the decay phase of many bursts is due to rp processing
and that the difference between PRE and non-PRE bursts is due to a
difference in H abundance in the burning layer. The difference between
the two studies, in addition to minor differences in the spectral
extraction and modeling we mentioned above, is that we quantitatively
attempt to distinguish between the cooling and rp burning. This is
only possible, however, if $\alpha$ may be considered constant. In
other words, studying the change in power-law decay index precludes
the study of the rp process, and vice versa. The exception to this
rule is for bursts for which it is more or less certain that there is
no H present in the burning layer, like in 4U
1820-30. \cite{kuuttila2017} found perhaps marginal evidence for
$\alpha$ evolution in that source, as judged from their Fig.~3, for
the phases where dynamical effects (PRE) are negligible. The
difference between soft and hard states is related to our finding of
the $S_{\rm Z}$ dependence in our study (Fig.~\ref{figzp}).
  
It may be questioned whether the details of the burst profile are only
due to processes within the neutron star or if the burst signal is
influenced by the accretion flow. There have been increasing reports
of the latter
\citep{zan11,chen2012,deg13,zand2013,ji2014a,ji2014b,worpel2013,worpel2015},
so this is a genuine concern. Fortunately, our attempt to take this
into account, through the $f_{\rm a}$ factor, does not make much
difference in the results on the distributions of fit parameters
(cf. Appendix~\ref{appb}).  With the assumption that contamination
from the accretion flow has been sufficiently taken into account in
our analysis, we discuss our findings on cooling and the rp process in
the following two subsections.

\subsection{Neutron star cooling}

The variation in the power-law decay index from burst to burst that we
find in this study in 501 bursts \typeout{NUMBERS}(Fig.~\ref{fig3}) is
similar to the variation found previously in a selection of 37
hydrogen-deficient bursts \citep{zand2014a}. This implies that the
cooling is largely independent of fuel composition or reaction
chain. This in turn implies that the variation in the heat capacity of
the ashes is small. Furthermore, we find that the broadness of the
distribution varies from source to source, with some having a much
narrower distribution and others having a distribution almost just as
broad as the whole sample, see Fig.~\ref{figpowc}. It is noticeable
that the narrower distributions occur together with rather constant
accretion fluxes.  This may be related to the fact that the burning
regime (see Sect. \ref{sec:intro}) and ignition depth change when the
accretion flux changes. It is also notable that the average index
differs from source to source. This is particularly interesting for
the three narrow distributions shown in Fig.~\ref{figpowc}, peaking at
$\alpha=1.62$, 1.91, and 2.06. This may again be due to systematically
different ignition depths. Unfortunately, there is no robust
model-independent method to determine the ignition depth for all our
bursts to further investigate this dependency.

In Fig.~\ref{fig8} a clear dependence is visible between decay index
and power-law duration. The bursts cluster on two branches that each
show a decreasing trend of decay index with power-law duration.
Figure~\ref{fig8} is a confirmation as well as an elaboration of a
tentative finding in Fig.~7 of \cite{zand2014a}. The duration of a
burst is rather directly dependent on the ignition depth. The shortest
bursts have ignition column depths of about 10$^8$~g~cm$^{-2}$. The
picture drawn by Fig.~\ref{fig8} is in rough agreement with
theoretical expectations \citep[see Fig.~6 in][]{zand2014a} that the
highest values for $\alpha$ are found in the earliest phases of bursts
(i.e., when the burst flux is above 10\% of the Eddington limit) with
the smallest ignition depth (therefore shortest durations). The reason
is that the photons dominate the electrons and ions at shallow
depths. We note that we fit only the latter part of the decay, when
the flux is half of the peak flux and possibly lower for bursts where
part of the flux goes into kinetic energy of a wind. This excludes the
high-flux regions \citep[see Fig.~6 in][]{zand2014a} where $\alpha$ is
steepest and is expected to change the fastest. This may explain
partly why our analysis is not sensitive to changes in $\alpha$.
Almost all UCXB, hydrogen-deficient, bursts are on the short
branch. Hydrogen-rich bursts spread over both branches. Those that are
on the short branch probably burn predominantly helium (in the
pure-helium burning regime, see Sect.~\ref{sec:intro}). This shows
that hydrogen plays an essential role in generating flat sub-Eddington
peaks. We note that if we consider only the blue data points from 4U
1636-536 in Fig.~\ref{fig8}, the $\alpha$ versus duration dependence
is opposite to that of the general trend. This explains why the
diagrams of $\alpha$ and $f$ with $S_{\rm Z}$ or pre-burst flux look
counter-intuitive. The reason is that there are two types of bursts
emanating from 4U 1636-536: hydrogen-rich and hydrogen-poor bursts. In
hydrogen-rich bursts, more photons are produced per gram and the
balance favors photons over ions and electrons, and the decay index
increases.

The picture drawn here of the power-law component is somewhat
confusing. On the one hand, it is unexpected that a single power law
describes the cooling of many bursts. This was already clear from the
previous study \citep{zand2014a}. It may be related to the dominance
in the heat capacity of one species of particles (ions, electrons, or
photons). On the other hand, if a single power-law decay index
describes the cooling of neutron stars, why does it differ from burst
to burst?  The diversity of power-law decay indices over all bursts,
also from a single neutron star, seems related to the diversity in
accretion flow, see for instance Fig.~\ref{figzp}, and as a result, on
the diversity of the ignition depth (see above). Possibly, both hands
can be joined: the non-detection of decay-index change in single
bursts is replaced by a detection of a variation in decay index from
burst to burst from the same neutron star.

\subsection{rp process}

About half of all bursts\typeout{NUMBERS} need the addition of the
Gaussian, the others are sufficiently well described by only a power
law over a dynamic flux range of, in general, 10$^2$. This is partly
due to a selection effect: for weaker bursts (intrinsically or because
of a relative large burster distance) or for strongly variable
accretion radiation, it is more difficult to detect the Gaussian.

In the assumption that the Gaussian is connected to the rp process,
its width $s$ is associated with the timescale of the rp process,
which has a direct correspondence to the extent of the nuclear chain
(see Fig.~\ref{figrp}).  The Gaussian width $s$ (Fig.~\ref{fig4})
shows a bimodal distribution: a sharp peak at 48 s with a spread of
about 10 s and a noisy continuum between 0 and roughly 50 s. Both
these components cover about half the population. These timescales are
as expected for the rp process, see Fig.~\ref{figrp}, if the nuclear
chain extends to at least about $^{42}$Ti. This is the first such
conclusion drawn from a large observational data study. The peak at 48
s is half due to bursts from GS~1826-24, which is rather stable (in
the 'hard state') in its behavior \citep{ube99,gal04,chenevez2016},
but the other half is due to other bursters.

The histogram of the Gaussian fluence fraction $f$ (Fig.~\ref{fig6})
is also bimodal, with a peak at 0.6 due mostly to the reproducible
bursts from GS~1826-24 and a continuum extending from 0 to about
0.4. The limited range suggests that there is generally less energy in
the rp process than in the $3\alpha$ burning. This fraction must be
related to the abundance of hydrogen and the extent of the rp
process. Burning H to Fe yields 5.6 MeV radiative energy per nucleon
\citep[where 35\% escapes through neutrinos as expected in the rp
  process, e.g.,][]{fuj87b}. Burning He to Fe yields 1.6 MeV per
nucleon. The observed fraction of $<0.65$ suggests that
$\frac{5.6X}{5.6X+1.6Y}<0.65$, so $\frac{X}{Y}<0.53$, if $X$ and $Y$
are the mass abundances of hydrogen and helium, respectively. This
compares to $\frac{X}{Y}=2.6-2.9$ for unprocessed fuel with
cosmological or solar abundances. During the Gaussian phase, hydrogen
therefore appears to be depleted by a factor of 5 in the burning layer
for all bursts from GS 1826-24 and some other bursts from other
sources, and at least by a factor of 13 for all other bursts.
Hydrogen depletion with respect to the canonical 70\% value is
probably due to hydrogen burning during the hot CNO cycle or during
the initial fast part of the burst (i.e., up to the first waiting
point in the rp chain).

The histograms of the Gaussian width $s$ and fluence fraction $f$ are
dominated by narrow components that are due to the textbook burster GS
1826-24. The burster also distinguishes itself by having the largest
and longest Gaussian component, rivaled perhaps only by the Rapid
Burster \citep{bag13,zand2017}. Other bursts have less hydrogen in the
burning layer.

4U 1636-536 is a nice test case for dependencies between burst
parameters and the accretion rate because the source has a
considerably variable accretion rate and exhibits frequent bursts at
many rates. Furthermore, RXTE has a large database on 4U
1636-536. There have been previous interesting studies on this data
set, concentrating on the burst oscillation behavior and burst
spectral evolution versus accretion rate \citep[e.g.,][]{zhang2013}.
We find an interesting dependency (Fig.~\ref{figz}) between the
Gaussian fluence fraction and the $S_{\rm z}$ parameter, which stands
for a proxy for the accretion rate. At the low end of the mass
accretion rate, we find a substantially larger Gaussian
component. This is consistent with the source moving to the so-called
burning regime 3 \citep[according to][see
  Sect.~\ref{sec:intro}]{fujimoto1981} where bursts are ignited in a
hydrogen-rich layer. \cite{zhang2013} find that there is a lack of
burst oscillations in this same burning regime.

Ninety-four percent of the bursts from UCXBs clearly lack a
significant Gaussian component, and 6\% show only a marginally
significant detection, which is consistent with the expectation that
UCXBs lack hydrogen and the Gaussian being representative of the rp
process.

\section{Conclusions}
\label{secconclusion}

Our simplified but successful approach to modeling the decay of
thermonuclear shell flashes on accreting neutron stars allows for the
following conclusions.

\begin{enumerate}
\item Almost all time profiles of X-ray burst bolometric flux can be
  better modeled by the combination of a power law and a one-sided
  Gaussian function than with an exponential function.
\item No dependency is found between the two light-curve components.
\item Most UCXBs lack significant Gaussian components, which is
  consistent with it being due to rp burning. The few exceptions are
  mostly from 4U 1728-34 and are marginal detections.
\item The decay index remains constant within each burst, at least for
  times when the flux is below 55\% of the peak flux.
\item There is no universal power-law decay index; it ranges for 80\%
  between 1.3 and 2.2. 
\item There is no unique decay index for any given neutron star, except
  perhaps for GS 1826-24.
\item The last two points are possibly connected to a spread
  in ignition depths.
\item The range of decay indices for H-rich systems is similar to that
  for H-poor systems;
\item GS 1826-24 is an exceptional burster in that it exhibits the largest
  rp component and the shallowest and most constant decay index.
\end{enumerate}

This investigation was made possible by the high throughput of the PCA
instrument on RXTE.  It may be verified with the similarly sensitive
LAXPC instrument on Astrosat \citep[e.g.,][]{agr06}, provided a
similar amount of bursts are observed. A most useful asset in this
regard would be an extension of the bandpass toward both lower and
higher photon energies and a larger detector effective area. New
missions to look forward to in this respect are the NICER mission
\citep[e.g.][]{gendreau2012,keek2016b} and the proposed mission
concepts LOFT \citep{feroci2012,zand2015,feroci2016}, eXTP
\citep{zhang2016} and Strobe-X \citep{ray2017}.

\begin{acknowledgements}
We thank Edward Brown and Andrew Cumming for useful discussions and an
anonymous referee for suggestions to improve the paper. This paper
uses preliminary analysis results from the Multi-INstrument Burst
ARchive (MINBAR\footnote{URL {\tt burst.sci.monash.edu/minbar}}),
which has received support from the Australian Academy of Science's
Scientific Visits to Europe program, the Australian Research Councilâ€
™sDiscovery Projects and Future Fellowship funding schemes and the
European Union's Horizon 2000 Programme under the AHEAD project (grant
agreement n. 654215).
\end{acknowledgements}

% \clearpage % inserted for referee version

\bibliographystyle{aa} \bibliography{references}

\appendix

\section{Is a Gaussian function a good representation of the rp process?}
\label{appa}

\begin{table}
\centering
\caption{Half-life and energy Q per $\beta^+$ decay. All half-lives
  and Q values were obtained from the REACLIB archive
  \citep{sakharuk2006}}
\label{tab2}
\begin{tabular}{lrr|lrr}
\hline\hline
Isotope & $t_{\frac{1}{2}}$ & Q (keV) & Isotope & $t_{\frac{1}{2}}$ & Q (keV)\\
\hline
$^{21}$Mg  &  5.658    &  5504.18  & $^{66}$Se  &  17.328   &  1818.97 \\
$^{22}$Mg  &  0.180    &  12860.3  & $^{67}$Se  &  6.931    &  4892.27 \\
$^{25}$Si  &  3.180    &  5514.43  & $^{68}$Se  &  0.007    &  8996.11 \\
$^{26}$Si  &  0.314    &  12816.8  & $^{71}$Kr  &  7.146    &  -213.75 \\
$^{29}$S   &  3.707    &  4399     & $^{74}$Sr  &  1.386    &  1707.62 \\
$^{30}$S   &  0.587    &  11050.04 & $^{75}$Sr  &  0.347    &  4314.16 \\
$^{33}$Ar  &  4.007    &  4662.76  & $^{76}$Sr  &  0.139    &  8745.81 \\
$^{34}$Ar  &  0.821    &  7564.61  & $^{79}$Zr  &  3.466    &  4449.44 \\
$^{36}$K   &  2.027    &  1857.63  & $^{80}$Zr  &  0.693    &  8011.69 \\
$^{37}$K   &  0.568    &  10904.68 & $^{83}$Mo  &  6.931    &  4136.41 \\
$^{39}$Ca  &  0.806    &  13181.45 & $^{84}$Mo  &  0.693    &  7063.91 \\
$^{42}$Ti  &  3.466    &  9181.46  & $^{87}$Ru  &  19.804   &  3813.97 \\
$^{45}$Cr  &  13.862   &  4883.08  & $^{88}$Ru  &  0.693    &  5083.941\\
$^{46}$Cr  &  2.666    &  6828.34  & $^{90}$Rh  &  41.258   &  4768.94 \\
$^{48}$Mn  &  23.104   &  6234.31  & $^{92}$Pd  &  0.693    &  5673.911\\
$^{50}$Fe  &  6.931    &  9091.04  & $^{95}$Cd  &  47.803   &  3292.97 \\
$^{53}$Ni  &  15.403   &  3854.93  & $^{96}$Cd  &  0.827    &  3320.97 \\
$^{54}$Ni  &  6.931    &  4615.01  & $^{97}$Cd  &  0.248    &  7827.497\\
$^{55}$Ni  &  3.667    &  10136.09 & $^{100}$Sn &  0.737    &  2679.14 \\
$^{58}$Zn  &  6.931    &  2887.37  & $^{101}$Sn &  0.231    &  3428.97 \\
$^{59}$Zn  &  3.301    &  7887.05  & $^{102}$Sn &  0.182    &  3568.97 \\
$^{62}$Ge  &  13.862   &  2238.67  & $^{103}$Sn &  0.098    &  4324.97 \\
$^{63}$Ge  &  3.466    &  4855.66  & $^{104}$Sn &  0.033    &  6224.911\\
\hline\hline
\end{tabular}
\end{table}

\begin{figure}[t]
\begin{center}
\includegraphics[width=\columnwidth]{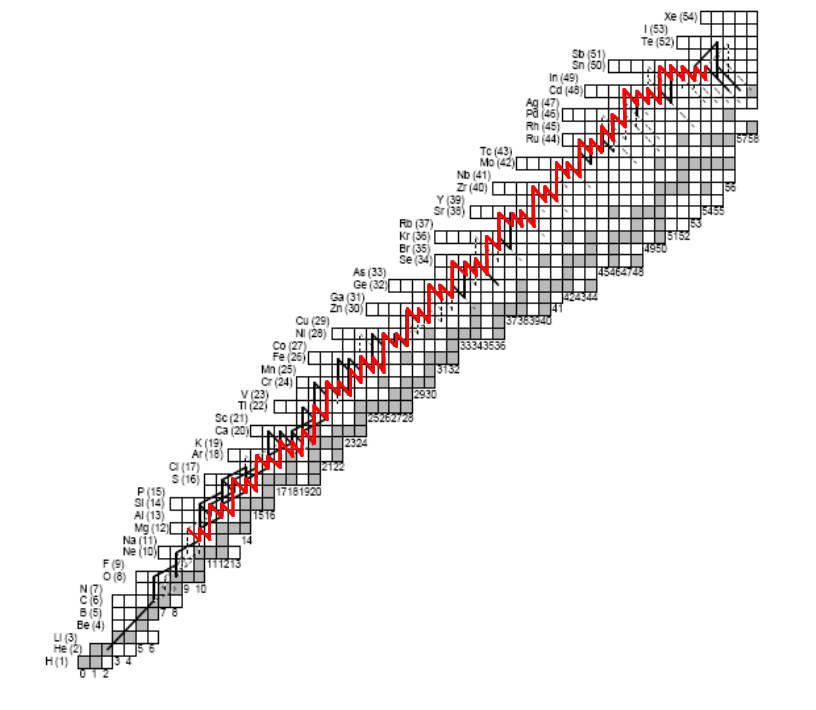}
\end{center}
\caption{Reaction chain followed in our simplified rp model in
  red, based on the reaction flows as prescribed by \cite{schatz2001}
  and \cite{wallace1981}.\label{fig7}}
\end{figure}

\begin{figure}[t]
\begin{center}
\includegraphics[width=\columnwidth,angle=0,trim=2.5cm 1cm 7.5cm
  16cm]{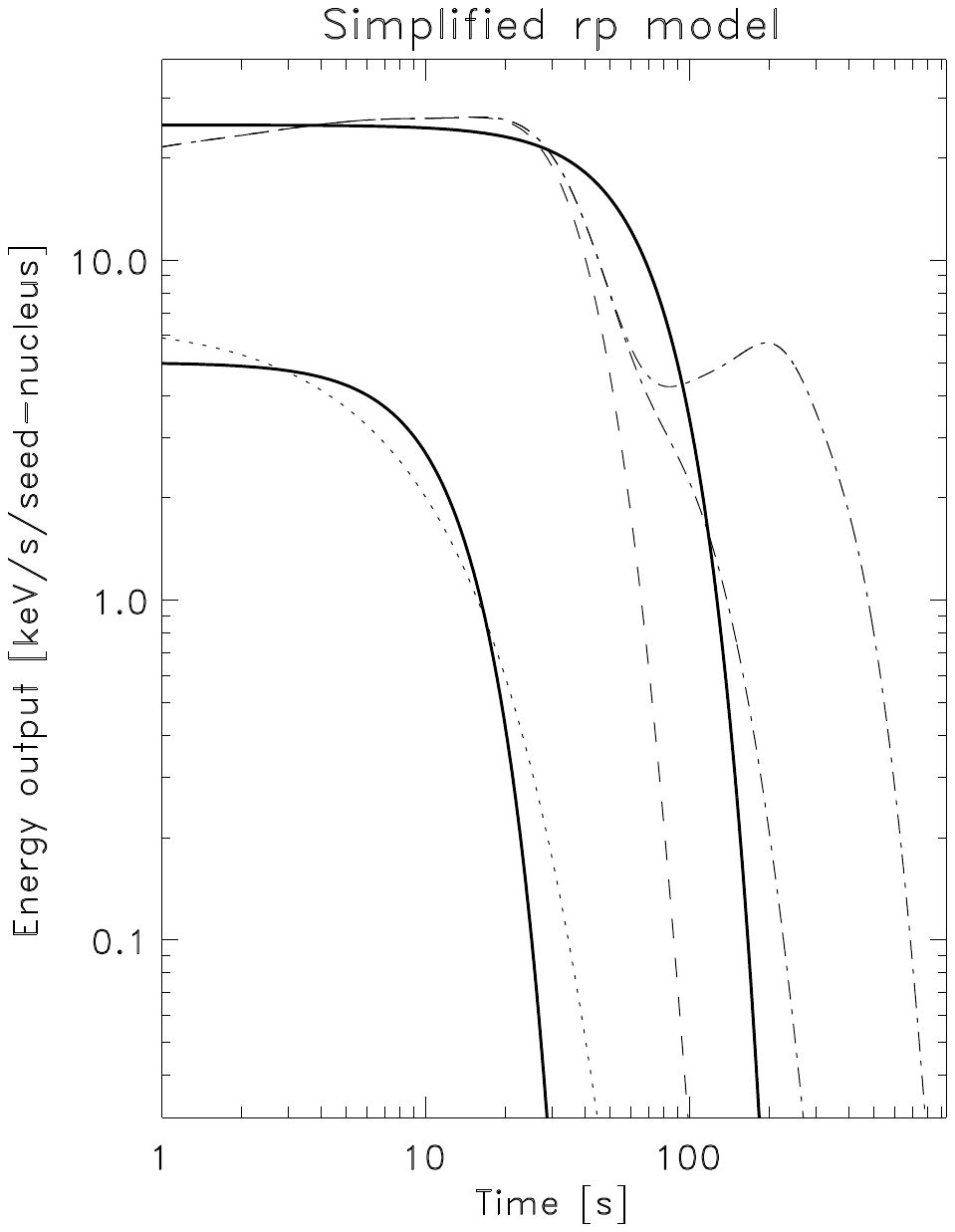}
\end{center}
\caption{Nuclear power according to our simplified rp process for
  end points at $^{21}$Mg (dotted curve), $^{42}$Ti (dashed),
  $^{53}$Ni and $^{104}$Sn (left dash-dotted and right dash-dotted). In
  these simplified calculations, $Q$ (see Table~\ref{tab2}) is up to 2.4
  MeV/nucleon for $^{53}$Ni and 2.6 MeV/nucleon for $^{104}$Sn. The
  lines are two examples of one-sided Gaussians with widths of 9 and
  50 s.\label{figrp}}
\end{figure}

\begin{figure}[t]  % from !t to t for referee version
\begin{center}
\includegraphics[width=\columnwidth,angle=0,trim=1.5cm 2cm 1.5cm 0cm]{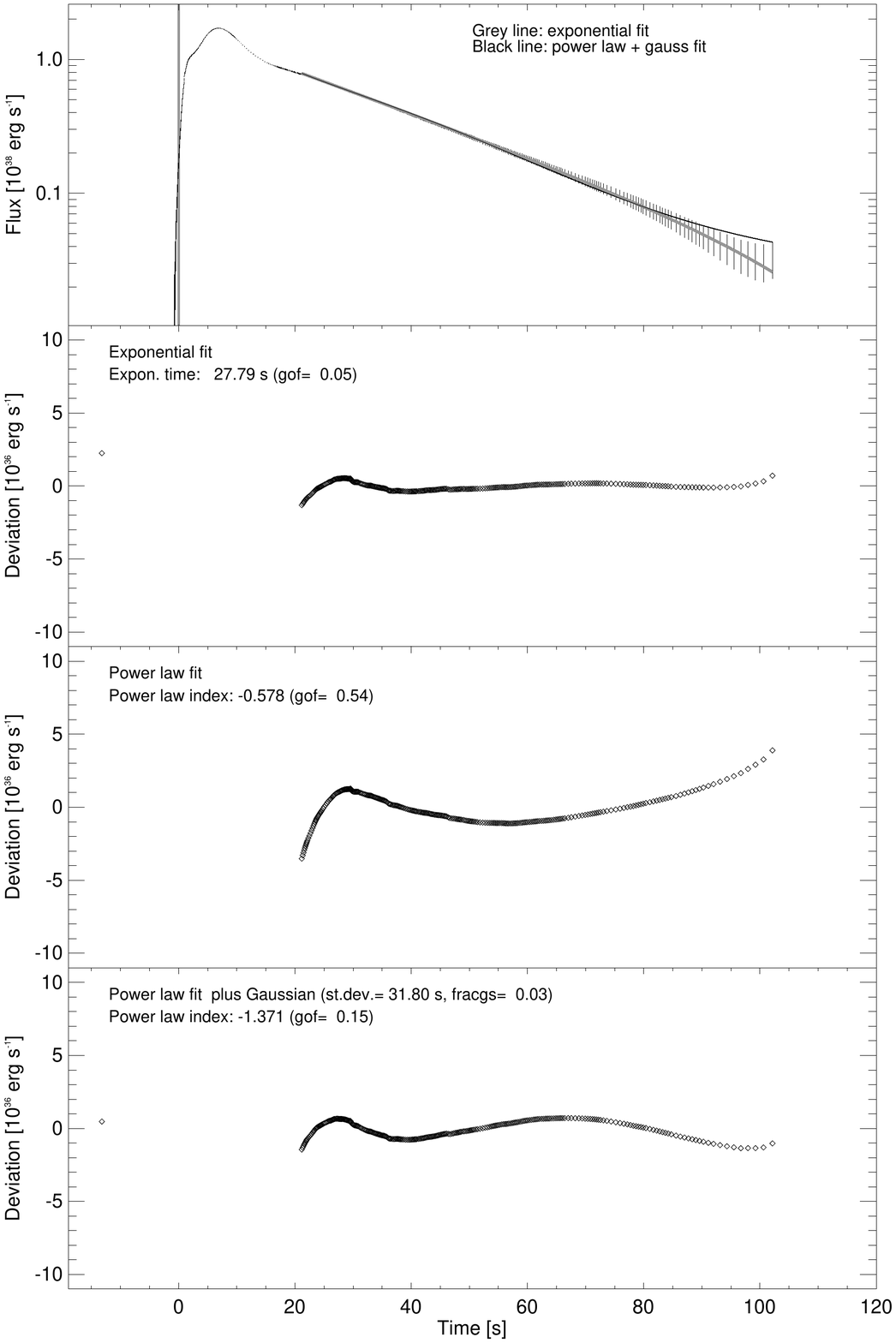}
\end{center}
\caption{Fit to KEPLER model 28 \citep[][]{lampe2016}, showing
  the effectiveness of the empirical model power-law plus Gaussian
  function. The model has a metallicity of 2\%, a hydrogen content of
  70.48\%, and an accretion rate of 8.2\% of Eddington, so it is expected
  to have a strong rp tail. Fiducial errors have been applied to
  calculate the goodness of fit (gof) with the reduced chi-squared
  formula. The gof numbers have no meaning in an absolute sense but
  may be compared with each other.\label{figkepler}}
\end{figure}

The one-sided Gaussian is a straightforward mathematical zeroth-order
description of the deviations of the bolometric flux time history from
power-law decays.  The question emerges whether a one-sided Gaussian
is a well-motivated zeroth-order representation of the rp process. We
investigated this in two ways.

\subsection{Simple nucleosynthesis model for the rp process}
\label{appa1}

We created a simplified model of the rp process, based on the reaction
chains as prescribed in \cite{wallace1981} and
\cite{schatz2001}. These reaction chains outline which proton captures
and beta decays can take place during the rp process. The reaction
chain we used was based on the work by Wallace and Woosley for
reaction chains with an end point up to $^{50}$Fe \citep[see
  Fig.~\ref{fig7} of this paper; for longer reaction chains,
  see][]{schatz2001}. The full reaction chain we used is presented as
the red line in Fig.~\ref{fig7}.  There are many points in the
reaction chain where both a proton capture and a beta decay are
possible. To avoid making our model too complex, at such particular
points we ignored the beta decay and assumed that the entirety of the
reaction would go through the proton capture. We also assumed a
constant high temperature and proton density so that proton capture
could be considered instantaneous and the timescales would depend
entirely on the beta decay half-lives if no subsequent proton capture
is likely (see Table~\ref{tab2}).  The simplified model consists of a
chain of equations, each calculating the abundance of a beta-decaying
isotope per 0.01 s time step. The abundance of each isotope depends on
the decay of the previous isotope and on its own decay.  We chose
$^{21}$Mg as the first isotope for our model because it is the first
isotope encountering beta decay when the reaction chain breaks out of
the CNO cycle for $T>3\times10^8$~K, and gave it a starting abundance
of 1. We calculated the power associated with each beta decay in
keV/s/seed-nucleus by adding the energy Q that is liberated with each
proton capture following the beta decay, and multiplying this energy
by the number of nuclei that decayed. Table~\ref{tab2} shows the
half-lives of the different isotopes in the reaction chain. We see
that there is one decay that has a negative Q associated with it,
$^{71}$Kr. This is because one of the proton captures following the
decay of a $^{71}$Kr atom, $^{72}$Kr(p,$\gamma$)$^{73}$Rb has a Q of
-7121.97 keV. This is the only proton capture in the full reaction
chain that has a negative Q. Because so many reactions with a positive
Q take place simultaneously, in a real burst there would likely be
enough energy available for this capture to take place nevertheless.
We convolved the power with a normalized power law with a decay index
of 1.6 to account for the fact that the energy released by the rp
process first has to travel through the atmosphere before it is
emitted by the photosphere and can be measured by our instruments.

Examples of resulting light curves for the simplified rp model are
given in Fig.~\ref{figrp}, together with two one-sided Gaussian
functions. One has to keep in mind that this figure shows only the
output of the rp process, not the output of the cooling after the
initial 3$\alpha$ flash. The rp curves show the increase in timescale
with the extent of the rp chain. The Gaussian functions are not very
good descriptors of the simplified rp process, particularly for the
secondary peak at about 100 s, but this peak is rather small and hard
to distinguish in the data (and, in fact, not seen in the most
sensitive data of a burst with a high rp content, see
Fig.~\ref{figtestcase}). Gaussian functions do cover the general
trends and provide a tool to quantify the observed
timescales. Therefore the timescale measured through the Gaussian
function in principle provides a measure for the chain length of the
rp process. Combining the timescale measurement with the amplitude
provides at least a verification of this chain length.

\subsection{KEPLER models}

\cite{lampe2016} published 465 simulations of thermonuclear X-ray
bursts with the KEPLER code \citep{weaver1978,woosley2004} for various
combinations of mass accretion rate, metallicity, and H abundance. The
KEPLER code includes an elaborate current nuclear network of 1300
isotopes that models CNO, triple-$\alpha$, $\alpha$p, and rp
burning. The compositional, temperature, and density changes are
tracked one-dimensionally along several zones in the radial direction.
Convection and thermohaline mixing are included through
one-dimensional prescriptions. Most of the bursts simulated by
\cite{lampe2016} are for accretion rates quite close to the Eddington
limit, while our observed bursts are usually far below this. We
focused on the 60 simulated bursts that have accretion rates below
10\% of the Eddington limit.

\cite{lampe2016} fit Eq.~\ref{eqn2} to their simulated light
curves. The decay indices they find are very broadly distributed,
between 0.4 and 7.6 for the subset of 60 bursts. Although most bursts
are H rich and formally fit unsatisfactorily with a sole power-law
function, the indices are much more diverse than we would expect. This
appears to be due to features in burst tails that are common in KEPLER
simulations but uncommon in observed bursts: flat shoulders that
extend up to 200 s and flat peaks extending up to 40 s followed by
decays with typical timescales a few times shorter. The latter may be
related to an incomplete calculation of radiation pressure effects
because all simulated bursts with flat peaks are typically
super-Eddington by a factor of 2, while the calculated photospheric
radius is not substantially increased. These features render a
detailed comparison with the model represented by Eq.~\ref{eqn3} less
useful. Nevertheless, as an illustration, we fit one simulated burst
(model A028) with our empirical model, see Fig.~\ref{figkepler}. In
this example, the power-law plus Gaussian function is a good fit and
the measured power-law index is in the range measured for most real
bursts. Peculiarly, the fit with an exponential function is slightly
better, while this is almost never the case for real data. This
testifies to the limited value of the KEPLER models as a benchmark for
our method.

\section{Results for a fixed $f_{\rm a}=1$ parameter}
\label{appb}

\begin{figure}[t]
\begin{center}
\includegraphics[width=0.90\columnwidth,angle=0,trim=1.5cm 7.5cm 1.5cm
  6cm]{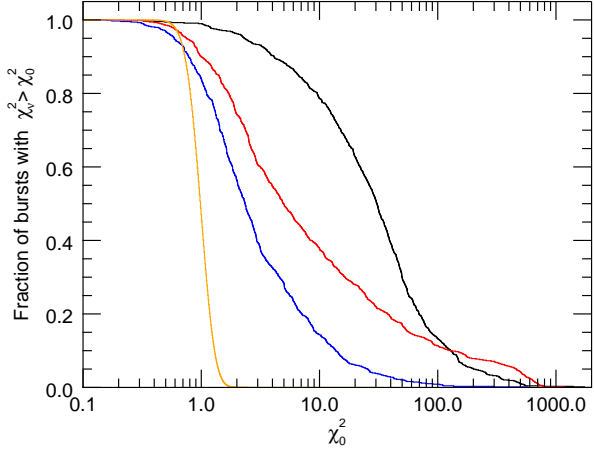}
\end{center}
\caption{Cumulative distribution of $\chi^2_\nu$ for fits with
  Eq.~\ref{eqn1} (black), \ref{eqn2} (red), and \ref{eqn3} (blue), for
  $f_{\rm a}=1$. The yellow curve shows the theoretical $\chi^2_\nu$
  distribution for 40 degrees of freedom (which is the
  average).\label{appfigb}}
\end{figure}

\begin{figure}[t]
\begin{center}
\includegraphics[width=0.90\columnwidth,angle=0,trim=1.5cm 8.cm 1.5cm 6cm]{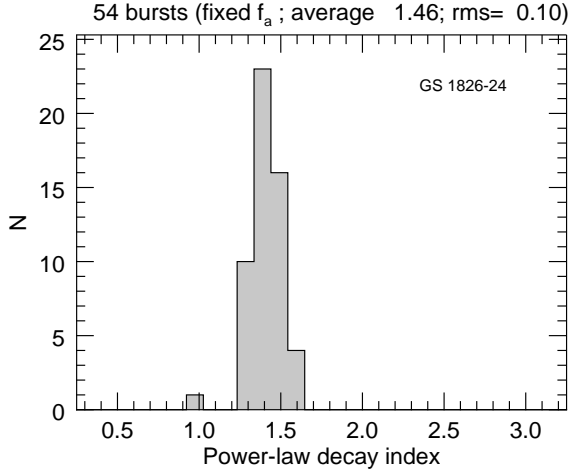}
\end{center}
\caption{Histograms of $\alpha$ for GS 1826-24 for $f_{\rm
    a}=1$.\label{appfiga}}
\end{figure}

\begin{figure}[t]
\begin{center}
\includegraphics[width=0.90\columnwidth,angle=0,trim=1.5cm 7.5cm 1.5cm 6cm]{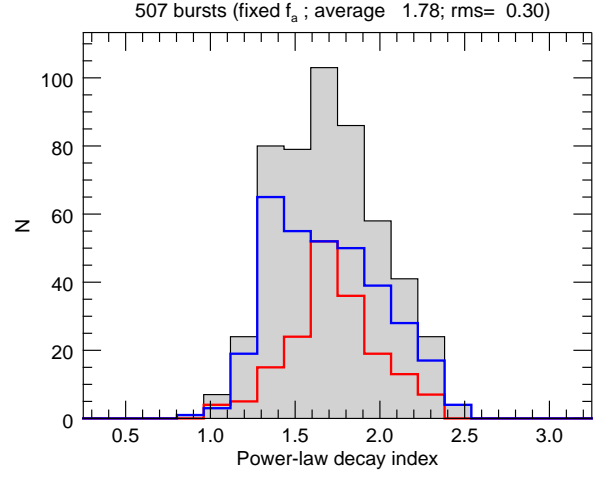}
\end{center}
\caption{Histograms of $\alpha$ for the $f_{\rm a}=1$ model. For a
  description, see the caption to Fig.~\ref{fig3}. \label{fig3app}}
\end{figure}

\begin{figure}[t]
\begin{center}
\includegraphics[width=0.90\columnwidth,angle=0,trim=1.5cm 7.5cm 1.5cm 6cm]{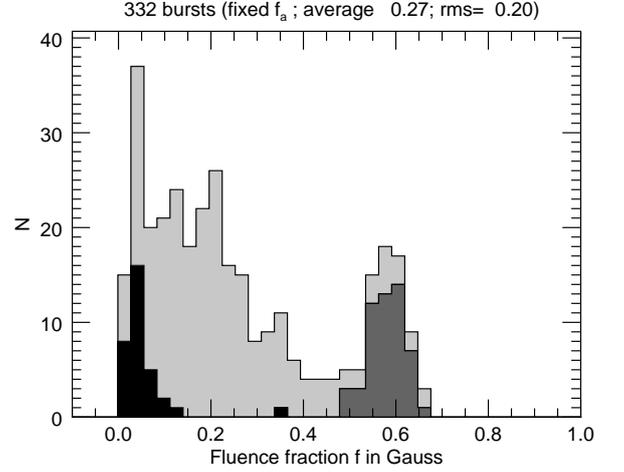}
\end{center}
\caption{Histogram of the fraction $f$ of the fluence contained in the
  Gaussian component for the $f_{\rm a}=1$ model. This may be
  compared to the histogram for the free $f_{\rm a}$ model in
  Fig.~\ref{fig6}.\label{fig6app}}
\end{figure}

We here present some illustrative results of a light-curve analysis
with $f_{\rm a}=1$. This entails a traditional X-ray burst analysis
where a certain pre-burst spectrum is subtracted from burst spectra,
and the assumption is that the accretion flow and emission is
unaffected by bursts.

Figure~\ref{appfigb} shows the cumulative $\chi^2_\nu$ distribution
for the fixed $f_{\rm a}$ model and is the equivalent to
Fig.~\ref{fig1} for the free $f_{\rm a}$ model. It shows that the fits
are substantially worse for a fixed $f_{\rm a}$ model. The main reason
is that statistical errors become larger with the additional $f_{\rm
  a}$ parameter. However, the conclusion is still valid that the power
law plus Gaussian best fits the data, followed by the power law and
the exponential function.

Figure~\ref{appfiga} shows the distribution of the power-law index
resulting from $f_{\rm a}=1$ for GS 1826-24. Of all sources in our
sample, GS 1826-24 has the narrowest distribution and is therefore a
good illustration of the effect of different choices for $f_{\rm
  a}$. For $f_{\rm a}=1$, the distribution is somewhat narrower (0.10
versus 0.12) and clusters at a shallower decay index (1.46 versus
1.62).  This shallower decay index may be explained as follows. The
index becomes steeper for free $f_{\rm a}$ solutions because the
low-flux data points carry less weight in the fit because of the
larger uncertainties. These low-flux points otherwise tend to pull the
power law to shallower decays.

Figure \ref{fig3app} shows the histogram of the power-law index for
the whole sample. This looks very similar to that for a free $f_{\rm
  a}$. The average is 1.78 versus 1.79 for a free $f_{\rm a}$ and the
rms is 0.30 versus 0.31\typeout{NUMBERS}.

Figure \ref{fig6app} shows the histogram of the fluence fraction of
the Gaussian. Again, it does not fundamentally differ from that for
the free $f_{\rm a}$ model.

\section{Online data}
\label{appc}

\renewcommand{\tabcolsep}{1.2mm} 

\begin{table*}[t]
\caption[]{List of all 1254 bursts included in our analysis. For a complete table, see online data \\ (see also URL {\tt http://www.sron.nl/$\sim$jeanz/nscoolrp/}.\label{tabj}}
\begin{center}
\begin{tabular}{lccccccrrrrrr}
\hline\hline
Source & MINBAR & MJD-start & Fit & Burst & Fit & Fit & \multicolumn{1}{|c}{$\alpha_2$} & $\chi^2_\nu$ & \multicolumn{1}{|c}{$\alpha_3$} & \multicolumn{1}{c}{$f$} & \multicolumn{1}{c}{$s$} & $\chi^2_\nu$ \\ 
       &   id   &           & type$^{\rm a}$ & start & start    & end  & \multicolumn{1}{|c}{}           &             &   \multicolumn{1}{|c}{}         &     & \multicolumn{1}{c}{(s)} &              \\
       &        &           &     & corr.$^{\rm b}$   &  time$^{\rm c}$     & time$^{\rm d}$  &   \multicolumn{1}{|c}{}         &             &    \multicolumn{1}{|c}{}        &     &   &              \\
       &        &           &     & (s)  &  (s)     & (s)  &   \multicolumn{1}{|c}{}         &             &    \multicolumn{1}{|c}{}        &     &   &              \\
\hline
4U 1608-52 & 2217  & 50164.69336& 2 & +1    &   9   &      & -1.796(020) & 3.13        & -1.824(010) & 0.021(004) & 37.5(6.5) & 3.10 \\
GS 1826-24 & 3393  & 53957.48029& 0 &       &       &      & -1.580(101) & 37.87       & -1.444(097) & 0.565(023) & 47.9(1.1) & 1.37 \\
           & 3481  & 54167.57565& 2 &       &       &      & -1.541(091) & 53.05       & -1.451(055) & 0.557(016) & 47.1(0.7) & 1.14\\
4U 1705-44 & 3871  & 55118.08693& 0 &       &       &      & -1.144(090) & 1.26        & -1.614(180) & 0.425(124)& 193.2(75.6) & 0.81\\
\hline\hline
\end{tabular}
\end{center}

\noindent
$^{\rm a}$The fit type is 0 when the burst was not included in
results, 1 when the single power-law result was included, and 2 when
the power law plus Gaussian was included. $^{\rm b}$Start time
correction in seconds with respect to the time given in
MJD-start. $^{\rm c}$Time delay after burst start time when fit
interval starts, if it is manually changed and not equal to the 55\%
criterion. $^{\rm d}$End time of fit in seconds since start time when
not equal to 300 s.
\end{table*}

Table~\ref{tabj} is an excerpt of a table that we provide online. It
lists for each of the 1254 bursts selected for this study the results
for models 2 and 3 (Eqs.~\ref{eqn2} and \ref{eqn3}, respectively),
including the goodness of fit $\chi^2_\nu$. Furthermore, it lists the
manual corrections that were made in incidental cases on the reference
time for the bursts, the start time of the burst, and the end time of
the burst. These three parameters are only listed when they were
different from the nominal values. All results are for a free $f_{\rm
  a}$.

\end{document}